\documentclass[twocolumn,twocolappendix]{aastex63}

\hypersetup{
 colorlinks  = true, 
 linkcolor  = red, 
 citecolor  = blue 
}

\usepackage{CJKutf8}
\usepackage{amsmath}
\usepackage{bm}

\newcommand\gj{\Gamma_{\rm j}}
\newcommand\betaj{\beta_{\rm j}}
\newcommand\betah{\beta_{\rm h}}
\newcommand\lj{L_{\rm j,39}}

\newcommand\rhoa{\rho_{\rm a}}
\newcommand\betaw{\beta_{\rm w}}
\newcommand\lw{L_{\rm w,39}}

\newcommand\Le{L_{\rm Edd}}
\newcommand\nam{n_{\rm a}}
\newcommand{\EpPev}{E_{p,\rm PeV}}
\newcommand{\ometil}{\tilde{\omega}}
\newcommand{\rtil}{\tilde{R}}

\received{December 21, 2024}
\revised{July 11, 2025}
\accepted{July 23, 2025}
\submitjournal{ApJL}

\shorttitle{Super Accretor as Super-PeVatrons }
\shortauthors{Wang et al.}

\begin{document}
\begin{CJK*}{UTF8}{gkai}

\title{Galactic Super-Accreting X-ray Binaries as Super-PeVatron Accelerators}

\correspondingauthor{Jieshuang Wang}
\email{jieshuang.wang@ipp.mpg.de}

\author[0000-0002-2662-6912]{Jieshuang Wang (王界双)}
\affiliation{Max Planck Institute for Plasma Physics, Boltzmannstra{\ss}e 2, D-85748 Garching, Germany}
\affiliation{Max-Planck-Institut f\"ur Kernphysik, Saupfercheckweg 1, D-69117 Heidelberg, Germany} 
\author[0000-0002-3778-1432]{Brian Reville}
\affiliation{Max-Planck-Institut f\"ur Kernphysik, Saupfercheckweg 1, D-69117 Heidelberg, Germany}
\author[0000-0003-1157-3915]{Felix A. Aharonian}
\affiliation{Max-Planck-Institut f\"ur Kernphysik, Saupfercheckweg 1, D-69117 Heidelberg, Germany}
\affiliation{Yerevan State University, Alek Manukyan St 1, Yerevan 0025, Armenia}
\affiliation{Dublin Institute for Advanced Studies, 31 Fitzwilliam Place, Dublin 2, Ireland}

\begin{abstract}
The extension of the cosmic-ray (CR) spectrum well beyond 1~PeV necessitates the existence of a population of accelerators in the Milky Way, which we refer to as Super PeVatrons. Identifying the nature of these sources remains a challenge to the paradigm of galactic CRs. Galactic super-accreting X-ray binaries, where the compact object accretes at a rate near or above the Eddington limit, can meet the energy requirement to supply the high-energy population of galactic CRs. We demonstrate that the trans-relativistic jets and/or winds of these powerful objects with kinetic energy luminosity exceeding $10^{39} \, \rm erg/s$, can accelerate protons to energies above several PeV. Detection of such super-accreting X-ray binaries through their ultra-high-energy $\gamma$-ray ``halos" and large-scale nebulae is also discussed.
\end{abstract}


\section{Introduction}

The locally measured spectrum of cosmic rays spans from sub-giga-electronvolts (GeV) to hundreds of exa-electronvolts (EeV). 
While CRs below the distinct spectral feature, called the {\it knee} around a few peta-electronvolts (PeV), are undoubtedly galactic in origin, 
it is believed that the galactic 
component of CRs extends well beyond these energies. 
The transition to extragalactic dominance can occur at $\approx 1 \, \rm EeV$, 
as supported by measurements of the dipole anisotropies and the mass compositions.
Observations reveal that the CR dipole direction 
lies close to the Galactic Center (GC) in the sub-EeV energy range, sweeping to larger angles above EeV energies \citep{Auger_anisotropy:2024}. 
A transition to extragalactic CRs close to EeV energies would require a rigidity around tens of PV for galactic CRs, even assuming the most extreme case in which iron dominates the end of the galactic CR flux.
A feature referred to as the {\it second knee} has been inferred at around 100 PeV, where recent measurements suggest in fact a dominance of C, N, O particles with a non-negligible fraction of protons \citep[e.g.,][]{IceTop_CR_proton:PhRvD:2019,TALE:ApJ:2021}.
Furthermore, the recent proton spectrum measured by LHAASO favors a spectral break rather than a cut-off above the {\it knee} \citep{TheLHAASOCollaboration:arXiv:2025}. 
These measurements necessitate consideration of what we call galactic super-PeVatrons,
which may provide the proposed second galactic CR component above several PeV to EeV energies \citep[e.g.][]{Axford:ApJS:1994,Thoudam:A&A:2016}.

In addition, the discovery of tens of ultra-high energy (UHE; $E \geq 0.1 \, \rm PeV$) gamma-ray sources by LHAASO \citep{LHAASO-catalog2} provides direct evidence for potential (super) PeVatrons candidates in the Milky Way. 
The gamma-ray spectra of large-scale diffuse sources, such as the Cygnus bubble \citep{LHAASO_CygBubble} and V4641 Sgr \citep{LHASSO_microquasar} extend to PeV energies without an obvious cut off in the spectrum. 
A hadronic origin of the emission has been suggested for the Cygnus bubble and some microquasars, which
would require these sources to accelerate particles to energies approaching 10~PeV. 
While a leptonic origin, due to competing radiative losses, would likewise indicate an extremely efficient super-PeVatron CR accelerator, as baryonic matter has been detected in the outflows of these sources \citep[e.g.,][]{Migliari:Sci:2002,Fabrika:ASPRv:2004}.
These findings reignite questions about the origin of CRs in the most challenging PeV-EeV energy range. In other words, the key issue becomes the maximum energy to which CRs can be accelerated in galactic sources. 
In this paper, we explore and discuss this question in the context of the most potent objects in the Milky Way - super-accreting X-ray binaries (XRBs), where the accretion rate is near or above the Eddington limit. 
Some of these sources are identified as Ultraluminous X-ray sources (ULX), which are defined as sources with X-ray luminosity above $10^{39}~{\rm erg/s}$. 
We here adopt the term super-accreting XRB to emphasize our focus on the kinetic power from XRB outflows.
Regarding the main focus and the objectives, our work, despite some overlap, differs from recent papers, e.g., by \cite{Peretti:A&A:2025}, who mainly discussed the ULX wind-driven nebulae in the context of diffusive shock acceleration, and by \cite{Ohira:arXiv:2024} who addressed the detection of extended $\gamma$-ray sources around microquasars in the context of CR propagation.

Supernova remnants (SNRs) have long been considered to be the principal contributors to galactic CRs \citep[e.g.][]{Ginzburg}, though from a theoretical perspective, acceleration of protons beyond 100 TeV presents a serious challenge \citep{Lagage:A&A:1983b}.
It is theorized that, with favorable magnetic field amplification conditions, SNRs can push acceleration to PeV energies \citep[e.g.][]{Zirakashvili:ApJ:2008,Bell:MNRAS:2013}. 
However, despite the detection of several young SNRs, 
unequivocal evidence for a hadronic origin of the TeV gamma-ray emission is lacking. 
Moreover, although inferred steep particle spectra at $\gg 1 \, \rm TeV$ seen in these sources 
cannot be considered as a decisive argument against SNRs \cite[see e.g., ][]{Malkov_steepspectra}, they present a non-negligible challenge \citep{Aharonian:NatAs:2019}, especially for multi-PeV energies.

Indirect evidence of proton acceleration to $>$ PeV energies can be inferred from the detection of a giant UHE gamma-ray bubble coincident with the stellar association Cygnus OB2 \citep{LHAASO_CygBubble}.
Observations of this, and other young ($<10$~Myr) stellar clusters at gamma rays above 0.1~TeV, such as W43 \citep{LHAASOW43} and Westerlund~1 \citep{HESS_WD1} have stimulated interest in the role of massive stellar clusters as PeVatrons \citep{Morlino:MNRAS:2021} or even super-PeVatrons \citep{Vieu:MNRAS:2022}. 
While a global theory for the galactic CR population using a stellar cluster/SNR model has been proposed \citep{Vieu:MNRAS:2023}, the steep spectra implied by observing stellar clusters/associations encourage consideration of alternatives.

The discovery of UHE gamma-ray emission from multiple galactic microquasars \citep{HAWC_V4641,LHASSO_microquasar}, has renewed interest in their potential to accelerate protons and other nuclei to multi-PeV energies.
These compact accreting binaries with mildly relativistic jets can generate hard gamma-ray spectra, as detected, for example, from the microquasars SS 433 and V4641 Sgr \citep{HESS_SS433,HAWC_V4641,LHASSO_microquasar}. 
These findings promote consideration of microquasars as potential contributors to the galactic CR flux above the {\it knee}. 
This hinges on the total and individual kinetic power of sources active over timescales of order the escape time for PeV CRs of about $\sim 0.1$ Myr. 
Here, we consider the subset of super-accreting XRB systems that operate in near- or super-Eddington states. 
The extreme luminosities observed from these sources, released through the kinetic energy of trans-relativistic outflows - winds and/or jets - can reach a level of $\gtrsim10^{39} \, \rm erg/s$ \citep{Jiang:ApJ:2014,King:NewAR:2023}. 
This power reflects a lower limit on the kinetic luminosity required to accelerate particles to rigidities of $10$\,PV.

In this paper, we propose that super-accreting XRBs are candidate super-PeVatrons that plausibly dominate the production of galactic CRs above the {\it knee}. 
In Section \ref{sec:PeVatron}, we introduce the general requirements for CR sources above the {\it knee}, and discuss potential sources.
The capability of super-accreting XRBs as super-PeVatron and CR sources is presented in Section \ref{sec:SAXRB}. 
The UHE emission from CR halos produced by super-accreting XRBs are discussed in Section \ref{sec:UHE-SAXRB}.
A summary is presented in Section \ref{sec:summary}.
In the Appendix, we present dynamics of wind- and jet-inflated nebulae.

\section{Requirements for Super-PeVatron candidates and their contribution to galactic cosmic rays}\label{sec:PeVatron}

We introduce the name ``super-PeVatrons" as cosmic-ray sources that accelerate particles from $>1$ PV to $100$ PV rigidities. 
While exploration of the physics of such objects is interesting in its own right, they have unavoidable implications for galactic CRs; namely these objects may explain the CR flux well above the {\it knee}, or even be responsible for a major fraction of galactic CRs more broadly, from GeV to multi-PeV energies. 
Below, we discuss the requirements addressing these questions in general terms.

\noindent
\subsection{Requirements to individual Super-PeVatrons}

The maximum kinetic energy a charge, $q=Ze$, can achieve is limited by the maximum electric potential difference across the system: $E_{\rm max} = Z e \bar{\mathcal{E}} R$, where $\bar{\mathcal{E}}$ is an effective electric field magnitude, which in the ideal magneto-hydrodynamic approximation is $\bar{\mathcal{E}}\sim \beta B$. The terms
$\beta$, $B$, and $R$ denote respectively the maximum velocity, magnetic field strength, and characteristic size of the acceleration zone \citep{Hillas:ARA&A:1984,Aharonian2002}. 
This energy limit may be cast in terms of the Poynting flux of the source, 
$L_B = {B^2\over 4\pi} \beta c\, A_{\rm eff}$,
where $A_{\rm eff}$ is the effective area through which the power flows.
We re-express the area as $A_{\rm eff}\equiv\ometil \pi \rtil^2$ for convenience. Here $\rtil = R_{\rm j}$ and $\ometil= 1$ for collimated jets, while for a quasi-spherical wind, $\rtil=R_{\rm w}$ and $\ometil = 4$ (see below, and Fig. \ref{fig:sketch}). 
The Poynting flux can be related to the kinetic power of the source: $L_{\rm K} = (\Gamma - 1) \rho c^2 \beta c\, A_{\rm eff} \equiv L_{\rm B}/\sigma$, where $\Gamma$ is the bulk fluid Lorentz factor, and $\sigma={B^2/ [4\pi (\Gamma - 1) \rho c^2]}$ is the magnetization parameter for a cold outflow. 
With these definitions, the maximum energy for mildly relativistic flows may be expressed as
\begin{equation}
  E_{\rm max} =35 Z \sigma_{-1}^{1/2}(L_{\rm K, 39}\beta)^{1/2} \ometil^{-1/2}~{\rm PeV}\label{eq:E_pmax},  
\end{equation}
where we adopt for convenience the shorthand $\xi_n \equiv \xi / 10^n $, in cgs units unless otherwise stated.
In this work, we do not consider the detailed physics of the acceleration process itself, but inspired by UHE observations of other extreme systems \citep[e.g.][]{LhaasoCollaboration:Sci:2021}, 
take the position that the accelerators operate at or near their maximum capability.
In this regard, our estimates should be interpreted as lower limits on the source requirements for a given $E_{\rm max}$.

The required kinetic luminosity has a strong dependence on $E_{\rm max}$, 
\begin{equation}
   L_{\rm K}\geq10^{38} \, (E_{\rm max}/10 \, \rm PeV)^2 \, \ometil 
   \, \beta^{-1} \sigma_{-1}^{-1} \, {\rm erg/s}
  \label{eq:power_Emax} .
\end{equation}
Thus, to achieve energies well beyond 10~PeV, even for the optimal configuration of the outflow with $\ometil \sim 1$ and $\beta \sim 0.1$ implying a mildly relativistic jet, and assuming a reasonably high magnetization ($\sigma \sim 0.1$), the kinetic luminosity should not fall below $10^{39} \, \rm{erg/s}$. 

\subsection{Super-PeVatrons as suppliers of Galactic CRs}

Eq.(\ref{eq:power_Emax}) represents a robust requirement for any ideal astrophysical outflow-driven source to operate as a super-PeVatron. 
Although being a super-PeVatron does not automatically imply a noticeable contribution to the galactic CR population, the significant power demand of Eq.(\ref{eq:power_Emax}) hints that a handful of super-PeVatrons alone can in principle supply galactic CRs in the region above the \emph{knee}, or even the entire galactic CR population. This will depend upon the fraction of the outflows' kinetic luminosity converted to CRs, the accelerated particles' spectrum, and the number of sources operating over the last $10^6-10^7$~years.   

The CR accelerators responsible for the local CR flux must replenish the energy lost via escape from the Galaxy to maintain a steady CR flux.
We adopt the assumption $\tau_{\rm esc}(E) = \tau_{10} (E/{\rm 10\,GeV})^{-\delta}$\,Myrs for the CR escape time from the Galaxy, where in the leaky-box model, $\tau_{10}$ is usually on the order of $100$, while $\delta$ lies in the range of $0.3-0.5$ \cite[see for example][]{Berezinskii}. 
Extrapolating this scaling to rigidities at the knee
with $\delta$ fixed, adopting parameters from \citep{Strong:ARNPS:2007}, one can estimate the power output in CRs as $L_{\rm CR}(E>E^*)\approx 1.5\times 10^{40} (\tau_{10}/100)^{-1} (E^*/{\rm 10\,GeV})^{-0.7+\delta}~{\rm erg/s}$.
For particles whose gyroradius exceeds the correlation scale of the local MHD turbulence, the scattering mean-free-path will increase as $E^{2}$. 
However, this does not imply an equivalent rapid shortening of escape times, due to the anisotropic nature of transport in the large-scale magnetic field of the Milky Way. For example, \citet{Giacinti2015PhRvD} have performed test particle simulations of CRs in different galactic magnetic field models with super-imposed Kolmogorov turbulence, and find that the grammage in the $0.1-100$ PeV CR energy range is consistent with energy-dependent escape times with index $\delta \approx 1/3$ breaking only above several PeV.
Allowing for some uncertainty in $\delta$, the required CR power from sources above the knee is $L(E>3\,{\rm PeV}) = A \times 10^{38}\,(\tau_{10}/100)^{-1}~{\rm erg/s}$, where $A\approx 1.4\,(12)$ for $\delta=1/3\, (1/2)$. 
Since the kinetic luminosity in the Super-PeVatron sources should be at least $10^{39} \, \rm erg/s$, a few Super-PeVatrons may be sufficient to explain the CR flux around and above the {\it knee} provided that approximately $10$ percent of the outflow's kinetic energy is converted to CRs. 
If the accelerated spectrum of super-PeVatrons below 1 PeV extends as a power-law with an index $\alpha \approx 2.7-\delta$, they could account for the entire CR population from GeV to tens of PeV energies, but the total power in CRs should satisfy $L_{\rm CR}(E>3\,{\rm GeV})\approx 0.9\times 10^{41} (\tau_{10}/100)^{-1} ~{\rm erg/s}$ for $\delta=1/2$. 
If the source spectrum flattens (i.e. $\alpha < 2.7-\delta$) at lower energies, super-PeVatrons contribute less below the {\it knee}. 
This is consistent with the usual paradigm, where SNRs play a dominant role \citep{Ginzburg}.

\subsection{Potential Candidates of Super-PeVatron}

\begin{figure}
  \centering
  \includegraphics[width=\linewidth]{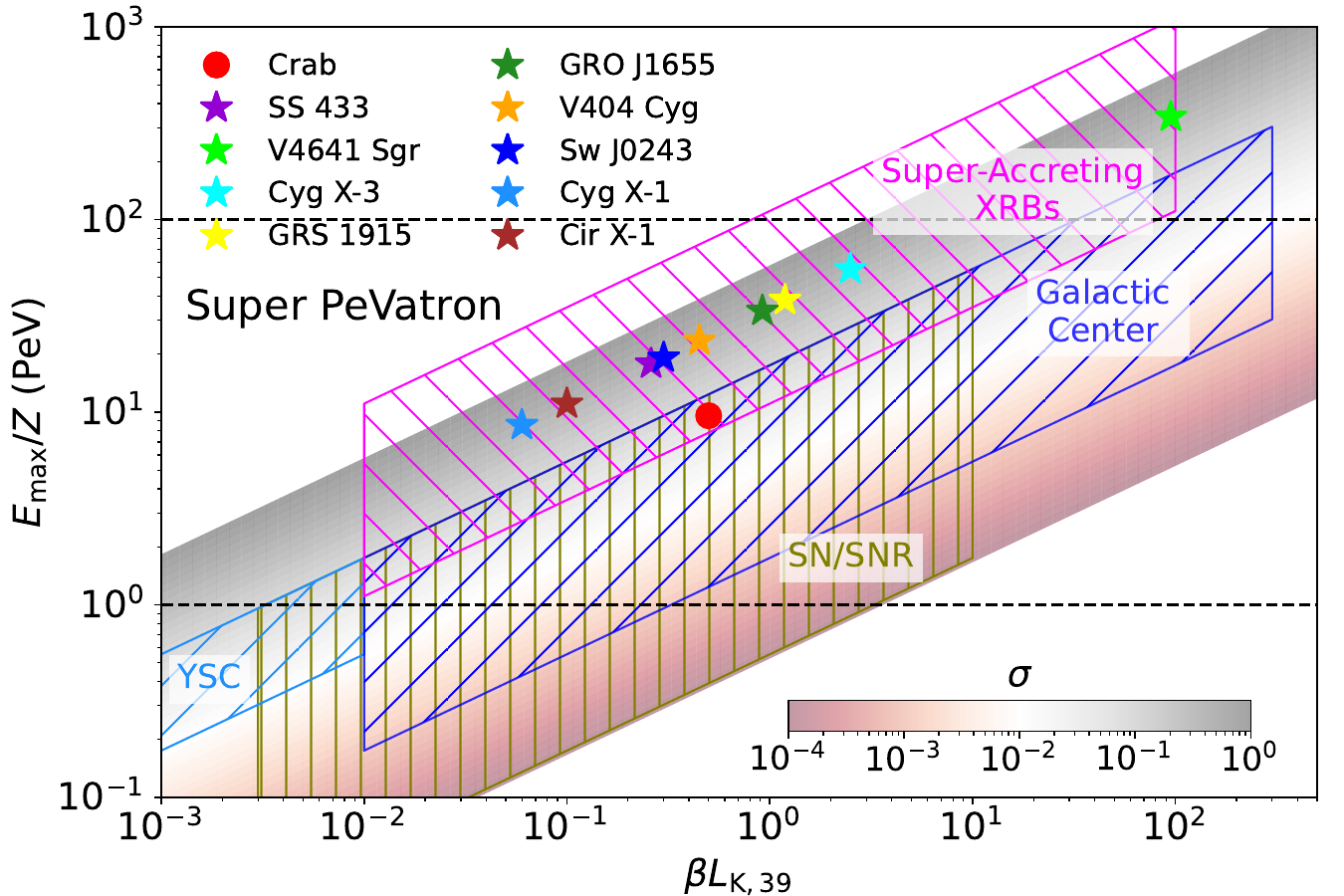}
  \caption{The maximum rigidity of Eq. (\ref{eq:E_pmax}) for young stellar clusters (YSCs), supernovae and their remnants (SNe/SNRs), pulsar wind nebulae (exemplified by Crab nebula), 
  the GC, and super-accreting XRBs.
  The corresponding kinetic power, magnetization, and velocity for this plot are summarized in Table \ref{tab:1}.
  We assume $\ometil=4$ for the background shading area with different values of $\sigma$. 
  For microquasar jets, the maximum energy can be higher than the shaded area because of $\ometil=1$.
  }
  \label{fig:hillas}
\end{figure}

Certain individual representatives of select source populations can satisfy the condition given by Eq.(\ref{eq:power_Emax}). The parameter space 
$[ \beta L_{\rm K}, E_{\rm max}/Z, \sigma]$ is shown in Fig. \ref{fig:hillas} for Supernova Remnants, Stellar Clusters, Pulsar Winds, the supermassive black hole (SMBH) - Sgr A* in the Galactic Center, and the objects of the main interest of this work - super-accreting XRBs characterized with (trans)relativistic outflows.  

\begin{table*}
  \centering
  \begin{tabular}{c|c|c|c|c}
  \hline
  Sources & Power ($10^{39}$~erg/s) & Velocity (c) & Magnetization & $E_{\rm max}/Z $\,(PeV)\\\hline
  YSC$^a$ & $0.1-1$ & $0.003-0.01$ & $10^{-4}-0.1$ & $0.1-2$ \\
  SN/SNR$^b$ & $0.1-10^3$ & $0.03-0.1$ & $10^{-4}-0.1$ & $0.03-55$\\
  Sgr A$^*$ $^c$ & $10-3\times10^3$ & $10^{-3}-0.1$ & $10^{-3}-0.1$ & $0.2-300$\\
  XRB$^d$ & $0.1-10^2$ & $0.1-1$ & $0.01-1$ & $1-10^3$\\\hline
  Crab$^e$ & $0.5$ & 1 & $0.06$ & 10\\\hline
  SS 433$^f$ & $1$ & $0.26$ & $0.1\sigma_{-1}$ & $18 \sigma_{-1}^{1/2}$\\
  V4641 Sgr$^g$ & $10^2$ & $0.95$ & $0.1\sigma_{-1}$ & $350\sigma_{-1}^{1/2}$\\
  Cyg X-3$^h$ & $5$ & $0.5$ & $0.1\sigma_{-1}$ & $55\sigma_{-1}^{1/2}$\\
  GRS 1915+105$^i$ & $1.7$ & $0.95$ & $0.1\sigma_{-1}$ & $31 \sigma_{-1}^{1/2}$\\
  GRO J1655-40$^j$ & $1$ & $0.92$ & $0.1\sigma_{-1}$ & $34 \sigma_{-1}^{1/2}$\\
  V404 Cyg$^k$ & $0.9$ & $0.5$ & $0.1\sigma_{-1}$ & $23 \sigma_{-1}^{1/2}$\\
  Swift J0243.6+6124$^l$ & $1.5$ & $0.2$ & $0.1\sigma_{-1}$ & $19 \sigma_{-1}^{1/2}$\\
  Cyg X-1$^m$ & $0.1$ & $ 0.6$ & $0.1\sigma_{-1}$ & $9 \sigma_{-1}^{1/2}$\\
  Cir X-1$^n$ & $0.2$ & $ 0.5$ & $0.1\sigma_{-1}$ & $11 \sigma_{-1}^{1/2}$\\
  \hline
  \end{tabular}
  \caption{The power, velocity and magnetization for different sources. 
  The upper panel is for source populations, while the lower is for individual sources. Values for the individual sources (excluding SS 433 and Cygnus X-3) correspond to extreme flaring states. 
  For values see 
  $a$: \cite{HESS_WD1,Vieu:MNRAS:2023};
  $b$: \cite{Zirakashvili:ApJ:2008, Bell:MNRAS:2013,Vieu:MNRAS:2022};
  $c$: \cite{Fermi_bubble,eROSITA_bubble};
  $d$: \cite{King:NewAR:2023} and references therein;
  $e$: 
  \cite{LhaasoCollaboration:Sci:2021};
  $f$: \cite{Fabrika:ASPRv:2004, Begelman:MNRAS:2006};
  $g$: \cite{Revnivtsev:A&A:2002,V4641_abundance:ApJ:2001};
  $h$:\cite{Veledina:NatAs:2024,Marti:A&A:2001,Miller-Jones:ApJ:2004};
  $i$: \cite{Fender:ARA&A:2004,GRS1915speed};
  $j$: \cite{Neilsen:ApJ:2016,Hjellming:Natur:1995};
  $k$: \cite{Tetarenko:MNRAS:2017,Zycki:MNRAS:1999};
  $l$: \cite{vandenEijnden:Natur:2018,vandenEijnden:MNRAS:2019};
  $m$: \cite{CygX1flare,CygX1flarejet};
  $n$: \cite{CirX1,CirX1jet}.
  }
  \label{tab:1}
\end{table*}

\subsubsection{Young stellar clusters}
Clusters of hot massive stars, with powerful winds, have received considerable attention in recent years. However, even the \emph{most powerful} known stellar cluster in the Milky Way, Westerlund 1, has a kinetic power $\lesssim 10^{39}$ erg\,s$^{-1}$ \citep{HESS_WD1}. Collective cluster winds, should they exist, must be super-Alfv\'enic i.e. $M_{\rm A} \gg 1$. Since in the non-relativistic limit $\sigma = 1/(2 M_{\rm A}^2) \ll 1$, clusters are not expected to act as super-PeVatrons \cite[see][for in depth discussion]{Vieu:MNRAS:2022}.  

\subsubsection{Supernovae (SNe) and their remnants (SNRs)} 

One can similarly estimate the power processed by the outer shock of a core-collapse SNR. 
The fast shock produced following the SN explosion will propagate into the wind of its progenitor. 
Assuming that prior to exploding the progenitor had a steady mass loss rate $\dot{M} = 4 \pi r^2 \rho v_{\rm wind}$, adopting numerical values typical for a red supergiant, the total power processed by a quasi-spherical shock is
\begin{align}
&L_{\rm K,SNR} 
\approx \int {1 \over 2} \rho u_{\rm sh}^3 A_{\rm eff} \approx {1 \over 2} \dot{M} u_{\rm sh}^2 (u_{\rm sh}/v_{\rm w}) \\
&\approx 10^{41}\frac{\dot{M}}{10^{-5} M_\odot / {\rm yr}} \left(\frac{u_{\rm sh}}{10^4 \,{\rm km/s}}\right)^3\left(\frac{v_{\rm wind}}
{30 \,{\rm km/s}}\right)^{-1}~{\rm erg/s} \, . \nonumber
\end{align}
Note that faster winds, such as those expected from Wolf-Rayet progenitors reduce the mass processing rate.
Despite the considerable power, in practice, the ambient magnetic field of an isolated core-collapse SNR is expected to be weak, since stellar winds must be super-Alfv\'enic, i.e. $v_{\rm A} \leq v_{\rm wind}$, and hence $\sigma \ll 1$. Current models of CR-driven magnetic field amplification, optimistically predict the maximum energy at $\gtrsim$ PeV \citep{Zirakashvili:ApJ:2008, Bell:MNRAS:2013}, though see \citet{Cristofari} for a critical view.
It is always possible that these models have overlooked some aspect of particle acceleration at SNR shocks, such as super-luminous SNe, hypernovae, or SNe in stellar clusters. 
It has been argued that turbulently amplified magnetic fields in the cores of compact massive stellar clusters may enhance the maximum energy in SNRs from dead massive stars therein \citep{Vieu:MNRAS:2022}.

\subsubsection{Pulsar Wind Nebulae} 

The detection of PeV photons from the Crab Nebula confirms that powerful pulsars can accelerate particles to multi-PeV energies 
\citep{LhaasoCollaboration:Sci:2021}. 
However, based on the current pulsar catalogue \citep{ATNFcatalog,Fermi_pulsar_2023}\footnote{https://www.atnf.csiro.au/people/pulsar/psrcat/}, the total spin-down power of known pulsars is around $3.6\times10^{39}~{\rm erg/s}$ with 4 sources having $\sim10^{38}~{\rm erg/s} $. The hadronic fraction of this power is unknown and the matter content may in fact be dominated by electron/positron pairs, making their contribution to the galactic CR population above the knee unclear.
We nevertheless include the Crab as an example in Fig. \ref{fig:hillas} and Table. \ref{tab:1}.

\subsubsection{GC and Galactic Outflows} 
The presence of a PeVatron source in the GC region was reported based on gamma-ray observations, the CR acceleration being tentatively related to past activity of the GC, especially the SMBH Sgr A$^*$ \citep{HESS_GC}.
It has furthermore been suggested that past Seyfert-like activity of Sgr A$^*$ or the past starburst activity produced the Fermi and eRosita bubbles \citep{Fermi_bubble,eROSITA_bubble}.
Therefore, although the suggested velocity for these large-scale bubbles is $\mathcal{O}(10^3)$~km/s, outflows could be launched at a higher velocity at the GC. 
In this case, the GC could conceivably be a super-PeVatron, as shown in Fig. \ref{fig:hillas} and Table. \ref{tab:1}. 
However, since the GC has been in a quiescent state in the recent past, its contribution to the current CR flux above knee remains to be studied.
CR acceleration has also been suggested to occur at the CR-driven galactic wind termination shock, \citep[e.g.,][]{Volk:A&A:2004}, though \citet{Mukhopadhyay:ApJ:2023} argue that it cannot fully account for the CR spectrum above PeV energies.
In the following sections, we restrict our focus to super-accreting XRBs as potential super-PeVatron CR sources.

\section{Super-accreting XRBs as super-PeVatron and CR sources}\label{sec:SAXRB}

Many galactic XRBs are found to be in near- or super-Eddington accretion phase intrinsically or transiently. 
The possible contribution of such systems to galactic CRs has been considered previously \cite[e.g.,][]{Heinz:A&A:2002,Cooper:MNRAS:2020}.
In table \ref{tab:1}, we list these super-accreting XRBs with jetted activities, i.e. microquasars. 
For most sources, the luminosity is taken from the extreme flaring state.
We also provide the corresponding maximum energies based on the Hillas criterion, since microquasars have been observed to be highly effective accelerators. 
For example, it was found that to account for the detected TeV emission from SS 433, acceleration needs to operate close to the maximum efficiency \citep{HESS_SS433}.
For super-accreting XRBs identified in X-rays, we further assume that the kinetic power of the outflows is comparable to the X-ray luminosity, since the X-ray activities reflect the accretion rate and state. 
Trans-relativistic outflows, via jets and/or winds, can be launched during the accretion process in these sources, with power of order the Eddington luminosity, $L_{\rm w} \gtrsim L_{\rm Edd} = 10^{39} (M/10M_\odot) \, {\rm erg/s}$. 
Observations suggest that these outflows can have velocities $\beta \gtrsim 0.1$.
Additionally, these outflows are expected to carry significant amounts of baryonic matter \citep[e.g.][]{Migliari:Sci:2002,vandenEijnden:Natur:2018}.
Thus, as shown in Fig. \ref{fig:hillas} and Table. \ref{tab:1}, in principle these sources can accelerate particles to 100~PV rigidities. 
Note for transient sources, one needs to take the duty cycle into account for calculation of their contribution to CR flux.

Apart from X-ray observations, super-accreting XRBs can also be inferred from their nebulae and large-scale halos as revealed by recent detections of XRBs in TeV-to-PeV gamma rays. 
In particular, the diffusive halos by CR sources provide important information on their recent past activities. 
Thus, the number of super-accreting XRBs could be revealed by future observations. 
We summarize the main features of the nebulae in Appendix \ref{sec:SAXRB-dyanmics}, in which we show that the nebula size is mainly determined by the kinetic power and age of the outflow.
In the next section, we discuss the UHE emission of these sources. 
In the following we take V4641 Sgr as an example.

A bolometric luminosity $\approx10^{41}~{\rm erg/s}$ was reported during an X-ray outburst for V4641 Sgr \citep{Revnivtsev:A&A:2002}. 
The jet velocity was inferred to be $\beta_{\rm j} \approx 0.99$ \citep{V4641_abundance:ApJ:2001}.
The corresponding maximum energy limit is $E_{\rm max,V4641}\approx 350Z\sigma_{-1}^{1/2}$~PeV, much higher than the maximum photon energies currently detected \citep{LHASSO_microquasar}. 
To explain the gamma-ray emission with hadronic interactions, the total energy in relativistic protons needs to be $\sim10^{50}$~erg \citep{HAWC_V4641}, which would require the source to have been in a high state for an extended period. 
This active period can be approximately estimated, assuming that its gamma-ray morphology is dominated by the nebula. 
This depends sensitively on its type, i.e., jet- or wind-driven.
For a jet-driven nebula with $\beta_{\rm j} \approx 0.99$, 
an active time of $t=50 n_a^{1/3} L_{\rm j,39}^{-1/3}$~kyr and $t_{\rm dec,0}=0.004L_{\rm j,39}^{2/3} n_a^{-2/3} $ (Eqs. \ref{eq:jet_Z} and \ref{eq:cocoon_R}) would be enough for $Z_{\rm j}=100$~pc and $R_{\rm c}=20$~pc for V4641 Sgr \citep{HAWC_V4641}. 
While for a wind-driven nebula (Eq. \ref{eq:R_FS}) with $R_{\rm w,FS}\approx100$\,pc, an age of $t=0.8n_a^{1/3}\lw^{-1/3}$\,Myr is required. 
Note these estimates assume that the outflow is steady. The ages are larger for variable sources.

To address their global contribution to CRs above the knee, we need to understand the prevalence of super-accreting XRBs in the Milky Way, which will depend on the star-formation rate.
Here we adopt the results of population synthesis studies on ULXs \citep{Wiktorowicz:ApJ:2017,Shao:ApJ:2020}, which usually operates at the near- or super-Eddington accretion rate persistently. 
It has been demonstrated that a Milky Way-like galaxy can maintain a population of $N \sim 10$ for a constant star formation rate ($6M_\odot/\rm yr$). 
This number may increase to $N \sim 10^2$ in the case of strong starburst activity. 
We here assume a fraction ($\epsilon_{\rm CR}$) of kinetic power of XRB outflows can be transferred to CRs, and that the kinetic power of the XRB outflows follows the X-ray luminosity function of ULXs.
We adopt a luminosity function with $dN/dL\propto L^{-1.6}$ and a cutoff at $10^{41}~{\rm erg~s^{-1}}$ based on the observation of high-mass XRBs in nearby galaxies \citep{Mineo:MNRAS:2012,King:NewAR:2023}.
The average kinetic power can be estimated as $\bar{L}_{\rm XRB}=\int_{10^{39}}^{\infty} L\,dN /\int_{10^{39}}^{\infty} dN=8\times 10^{39}~{\rm erg/s}$.
In addition, as super-accreting XRBs can be variable, we introduce a factor ($\epsilon_{\rm DC}$) to account for the duty cycle of the strong outflow state.
Note this duty cycle may differ from that of X-rays, 
since X-rays are subject to absorption and may not fully trace the kinetic power.
Additionally, X-rays only represent the current activity, and sources may have been more active in their recent past, as must have been the case for V4641 Sgr \cite{HAWC_V4641,LHASSO_microquasar}. Thus we propose XRB nebulae (Appendix) and XRB halos (Section \ref{sec:UHE-SAXRB}) as complementary probes to trace their activity.
The global CR injection flux by $N$ sources is then $L_{\rm CR}=\epsilon_{\rm CR}\epsilon_{\rm DC}N \bar{L}_{\rm XRB}$.
As indicated by the observations \citep{HESS_SS433}, we assume an acceleration spectral indices $s=2$ for energies from 3\,GeV to above 3\,PeV, consequently, a population of around $N\approx(2.5-21)\epsilon_{\rm CR,-1}^{-1}  \epsilon_{\rm DC}^{-1}(\tau_{10}/100)^{-1} $ sources for $\delta=1/3-1/2$ can meet the observed CR flux above the knee.

CRs produced by super-accreting XRBs could have super-solar abundances of elements.
For example, X-ray observations suggest a factor of 2-10 times higher than solar abundance of heavy elements, including S, Si and Ni, in the jet of SS 433 \citep{Fabrika:ASPRv:2004,SS433_abundance:A&A:2005,SS433_abundance:AstL:2018}; and it is also found that V4641 Sgr contains a massive B-star, whose $\alpha$-process elements, including N, O, Ca, Mg, and Ti, can be 2-10 times higher than solar-abundance levels from optical spectroscopic observations \citep{V4641_abundance:ApJ:2001}. 
Such features can be tested through future CR mass-composition measurements at sub-EeV energies.

\section{UHE emission of hadronic halos}\label{sec:UHE-SAXRB}

We consider here only hadronic processes, since we focus on UHE emission of gamma rays or neutrinos as tracers of $>10$~PV CR production.
At such extreme energies leptonic scenarios are unavoidably limited by cooling.
In the typical interstellar magnetic field of $\sim 5\,\mu$G \citep{Beck:A&ARv:2015}, synchrotron cooling dominates over inverse Compton losses on the CMB for PeV electrons. 
A strict upper limit to the distance that PeV electrons can travel from their prospective sources is $c\tau_{\rm c}\approx 100 (E_e/{\rm 1~ PeV})^{-1}$~pc considering both cooling mechanisms.
A much smaller halo size would be expected if the particle transport is dominated by diffusion.
Thus halos with size exceeding this scale at UHE energies will favor a hadronic origin.
Although hadronic interactions are also expected for heavier nuclei,
we focus on high-energy protons in the following section. 
The produced UHE gamma rays may be absorbed through $\gamma\gamma$ interactions during propagation. 
For photons at PeV energies, the absorption is mainly by interacting with the cosmic microwave background (CMB). 
The absorption probability on the CMB peaks at $\approx 3$\,PeV where the mean-free-path is $\approx8$\,kpc \citep{Gould66}.
Thus gamma rays above PeV energies at distances $\geq8$\,kpc will be significantly absorbed.
However, in this case, the existence of $10$-PV CRs can still be traced by PeV neutrinos and secondary emission from the absorbed PeV gamma rays.
For secondary electrons/positrons in the typical interstellar magnetic field, and their synchrotron radiation would peak at MeV gamma rays.
Below, we discuss UHE gamma rays from sources without correction for absorption.

Powerful outflows will inflate a nebula (see Appendix \ref{sec:SAXRB-dyanmics}). 
Particles can be accelerated in both the outflow or the magnetosphere of the compact star.
Here, we focus on the emission from the large-scale halo produced by escaped CRs (not to be confused with the nebula).
High-energy protons and atomic nuclei can produce emission via inelastic collisions (here termed $pp$ for simplicity) with ambient target gas.
The energy-loss time due to $pp$ interaction is $t_{pp}\approx 2\times10^7 n_{\rm t}^{-1}$~yr at PeV energies, where $n_{\rm t}$ is the target gas number density. 
This typically exceeds the dynamical time of the system, such that cooling is unimportant, with the maximum energy in principle limited only by the Hillas criterion described earlier.

\begin{figure*}
  \centering
  \includegraphics[width=\linewidth]{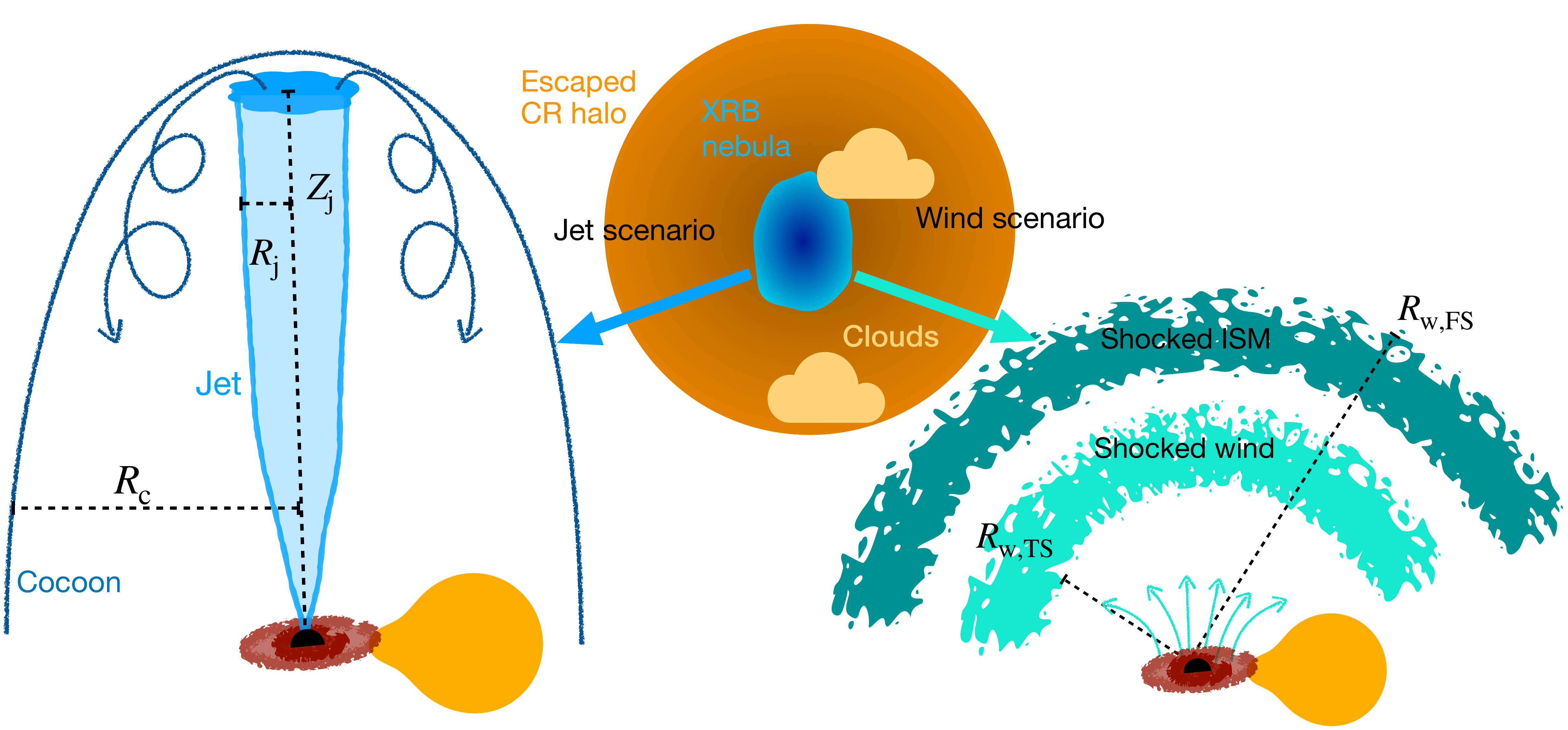}
  \caption{Illustration indicating the key features in the XRB nebula and halo (not to scale). 
  The nebula is inflated by the jet/wind.
  The CR halo is produced by escaping CRs interacting with the ISM.
  The jet may be collimated by the ambient medium. Close to the binary, the jet profile can be parabolic or conical. 
  Further away from the binary, it can be quasi-cylindrical.
  }
  \label{fig:sketch}
\end{figure*}

CRs escaping from the nebula into the ISM can fill a local volume, creating a diffuse source (or halo see Figure \ref{fig:sketch}).
Localized features can be produced in cases where dense clouds are present nearby the source, 
though here we consider an ISM with a uniform gas density.
We assume accelerated particles are distributed throughout the nebula, and leak out, diffusing isotropically in the surrounding ISM.
PeV protons diffuse to a distance $R_{\rm dif}\equiv \sqrt{4 D t}\approx120 D_{30}^{1/2} \EpPev^{\delta/2} t_{3}^{1/2}$~pc, where we have rewritten the diffusion coefficient as $D=D_{30} \EpPev^\delta$, with $ D_{30}=10^{30}~{\rm cm^2 s^{-1}}$ the diffusion coefficient at 1 PeV. This is approximately consistent with the previous estimate of escape time from the Galaxy. 
In the vicinity of several sources, a reduced diffusion coefficients has been reported \citep[e.g.][]{LHAASO_CygBubble}, leading to enhanced local confinement.
The CR energy density is $u(E_p,\Delta r,t)=\dot{Q}_{p}(E_p){\rm erfc}(\Delta r/R_{\rm dif}) /(4\pi D \Delta r)$ in the CR halo \citep{Atoyan:PhRvD:1995}, where $\dot{Q}_{p}$ is the CR injection rate, ${\rm erfc}$ is the complementary error function, and $\Delta r= r-R_S$, with $r$ being the distance to the central binary and $R_S$ the nebula size. 
The particle distribution follows approximately a profile $1/\Delta r$ at $\Delta r\lesssim R_{\rm dif}$.

The UHE gamma-ray flux from such a halo is $F_{\gamma}\approx uV/(4\pi d^2 t_{pp})\propto E^{2-s-\delta}$ for a power-law spectrum $\dot{Q}_{p}=\epsilon_{\rm CR} L_{\rm K} (E/3\,{\rm GeV})^{2-s}$, 
where the volume is $V=4\pi(r^3-R_S^3)/3$ and $d$ is the distance from Earth.
Since the diffusion length ($R_{\rm dif}$) is energy dependent, 
an energy dependent morphology is expected.
For a halo region, $R_{\rm Halo}> R_S$, we take $V\approx 4\pi R_{\rm Halo}^3/3$. 
The halo size can be further expressed as the angular size of the system, namely $\theta_{h}=R_{\rm Halo}/d=0.7^\circ D_{30}^{1/2} \EpPev^{\delta/2} t_{3}^{1/2} ({d/ \rm 10~kpc})^{-1} $.
The gamma-ray flux for $s=2\,(2.2)$ is
\begin{align}
   F_{\gamma}(&E_p={\rm 1\,PeV}) \approx 13\,(1)\times10^{-13}\epsilon_{\rm CR,-1} L_{\rm K,39}\\
  &\times{  (\theta/1^\circ)^2 \, n_{\rm t} D_{30}^{-1} }
   ~{\rm erg~cm^{-2}s^{-1}}.\nonumber
\end{align}
The large available power injected into CRs makes detection possible.
The resulting photons have energies extending up to $E_\gamma\approx0.1E_p$ below the cutoff.


In general, such super-accreting XRB halos are detectable with current or future gamma-ray telescope arrays, such as the Large High Altitude Air Shower Observatory \citep[][]{he18}, the Cherenkov Telescope Array Observatory \citep[][]{CTA2023APh}, or the Southern Wide-field Gamma-ray Observatory \citep[][]{SWGO2022icrc}. 
Neutrinos can also be produced with a comparable flux with typical energy $E_\nu\approx0.05E_p$, which could be detected by the next generation neutrino telescopes \citep[e.g.][]{Aiello:APh:2019,aartsen2021icecube,ye2023trident}.
Note such large-scale diffuse gamma-ray and neutrino sources are expected even after the super-accreting phase of the binary, as long as the high-energy particles are sufficiently well confined near their source.

\section{Summary}\label{sec:summary}
In this paper, we considered minimal requirements for super-PeVatron accelerators, emphasizing the role of galactic super-accreting XRBs as plausible candidates for the production of PeV-to-EeV CRs, i.e. the galactic population \emph{above} the knee. 
The extreme luminosity $\geq 10^{39}~{\rm erg/s}$ of super-accreting XRBs may facilitate acceleration of CRs beyond the {\it knee}, while around ten sources may be sufficient to account for the required flux, assuming that the kinetic power is comparable to X-ray flux.
The number of super-accreting XRBs can be revealed by detection of their UHE emission or their large-scale nebulae inflated by the jet or the wind.
In particular, although super-accreting activities can be episodic, the gamma-ray/neutrino halo can provide direct information on the total energy budget into CRs from their past activities.
The size and shape of the nebula, and possible UHE gamma-ray appearance is determined by the relative strengths of the wind and jet powers.
Thanks to their remarkable luminosity, UHE gamma-ray/neutrino emission produced by the accelerated CRs can be detectable by current and future observatories.

\acknowledgments
J.S.W. thanks Y. Shao for helpful discussions. 
We thank the referee for helpful comments.


\software{Astropy \citep{AstropyCollaborationApJ2022}; Matplotlib \citep{Matplotlib}; NumPy \citep{Numpy}
}

\appendix
\section{Super-Accreting-XRB outflow dynamics and Nebula size}\label{sec:SAXRB-dyanmics}

XRB nebulae have been used to infer super-accreting XRBs.
Multi-wavelength observations have shown that super-accreting XRB nebulae can be as large as tens to hundreds of parsecs, implying the kinetic power of the outflow must be on the order of $L_{\rm K} \sim (0.1 - 10) \times 10^{40}~{\rm erg/s}$ \citep[see Table 1 in][]{King:NewAR:2023}. 
The nebulae inflated by jets are expected to have a different morphology to that of wind-inflated nebulae, since winds are generally more isotropic, though a highly anisotropic wind could mimic a jet \citep{Churazov:A&A:2024}.
Here we assume the disk wind is quasi-isotropic. 
The typical dynamical features of jet/wind-inflated nebulae in the ISM are sketched in Figure \ref{fig:sketch}.
Below we discuss these two outflow types separately. 

\textbf{Jet nebulae:}
The dynamics of a jet propagating through an ambient medium has been well studied \cite[e.g.][see \cite{Marti:Galax:2019} for a recent review]{Begelman:ApJL:1989,Marti:ApJ:1997,Perucho:MNRAS:2007,Rossi:A&A:2008}. 
The basic features are shown in Figure \ref{fig:sketch}.
The jet head velocity ($\betah=\betaj w$) mainly depends on the jet-to-ambient density ratio $\zeta=\gj^2 n'_{\rm j}/\nam$, where $n'_{\rm j}$ is the jet number density in the proper frame, $\gj=(1-\betaj^2)^{-1/2}$ the jet bulk Lorentz factor, and $w(t)=\sqrt{\zeta}/(1+\sqrt{\zeta})$ for a uniform ISM \cite[e.g.][]{Marti:ApJ:1997}, $\nam =\rhoa/ m_p$ is the number density of the ambient medium, and $m_p$ is the proton mass.
Close to the central binary, the jet is dense with $\zeta\gg1$, 
which can move freely with a head velocity $\betah\approx\betaj$. 
The growth of the jet's radius, $R_{\rm j}(Z_{\rm j})$, with distance to the compact star depends on the external medium.
Far from the binary system, a slow jet head ($\zeta\leq1$) with a cocoon can form due to jet instabilities, entrainment of ambient materials, and/or the jet being intermittent.
We assume the deceleration take place a time $t_{\rm dec}$.
Hydrodynamic simulations show that the jet head velocity decays over time with approximately $w(t)= (t/t_{\rm dec})^{\alpha}$ with $\alpha=-1/3$ due to jet instabilities \citep{Scheck:MNRAS:2002,Perucho:MNRAS:2007}.
A fraction of jet energy $\eta_{\rm c} \approx 1-\betah/\betaj$ is deposited in a hot cocoon with a pressure $p_{\rm c}\approx \eta_{\rm c}L_{\rm j} t/V_{\rm c}$, where the cocoon volume is $V_{\rm c}\approx1.5\pi R_{\rm c}^2 \beta_{\rm h}c t$ and $R_{\rm c}$ is the cocoon radius.
The cocoon expansion speed can be estimated from the pressure balance $p_{\rm c}\approx\rhoa \dot{R}_{\rm c}^2$.
At late times with $\betah \propto t^{\alpha}$, we have $\dot{R}_{\rm c}\propto t^{-1/2 - \alpha/4}$ \citep{Scheck:MNRAS:2002,Perucho:MNRAS:2007}.
The original Begelman-Cioffi model assumes $\alpha=0$ \citep{Begelman:ApJL:1989}, while $\alpha=-2/5$ is used for the self-similar solution \citep{Komissarov:MNRAS:1998}.
Adopting here $\alpha=-1/3$ \citep{Scheck:MNRAS:2002},
the jet length and cocoon radius are
\begin{eqnarray}
  Z_{\rm j}&=\int \betah c {\rm d}t \approx 
  46\betaj t_{\rm dec,0}^{1/3}t_3^{2/3} ~{\rm pc};\label{eq:jet_Z}\\
  R_{\rm c}&=\int\dot{R}_{\rm c}{\rm d}t \approx 1.3{\lj^{1/4}\eta_{\rm c}^{1/4}t_3^{7/12} \over t_{\rm dec,0}^{1/12}n_a^{1/4}\betaj^{1/4}}~{\rm pc},\label{eq:cocoon_R}
\end{eqnarray}
where $t=t_3$~kyr, and $t_{\rm dec}=t_{\rm dec,0}$\,yr.
The cocoon expands at a velocity $\dot{R}_{\rm c}\approx10^3{\lj^{1/4}\eta_{\rm c}^{1/4}t_3^{-5/12} t_{\rm dec,0}^{-1/12}n_a^{-1/4}\betaj^{-1/4}}~{\rm km/s}.$
Particles can be accelerated inside the jet and/or at the recollimation/termination shocks, and transported into the cocoon.
The nebulae powered by jets can be elongated and can have a size of tens of parsecs.
However, a spherical-like ($Z_{\rm j}=R_{\rm c}$) jet-cocoon system is possible if the jet decelerates appreciably to a velocity $\betaj \approx 0.1 \lj^{1/5}\eta_{\rm c}^{1/5}t_3^{-1/15} t_{\rm dec,0}^{-1/3}n_a^{-1/5}$. 

\textbf{Wind nebulae:} At near/super-Eddington accretion rates, mildly-relativistic disk winds, with velocity $ \beta_{\rm w}\approx0.1$, can be launched \citep[e.g.][]{Proga:ApJ:2000,Ohsuga:ApJ:2005,Jiang:ApJ:2014,King:NewAR:2023}. 
Far from the disk, the wind is quasi-spherical. 
The wind's luminosity can be close to the Eddington luminosity $L_{\rm w}\approx\Le=10^{39}(M/10M_\odot)~{\rm erg/s}$ \citep[e.g.][]{King:NewAR:2023}, where $M$ is the mass of the compact star. 
The system is characterised by a strong termination shock in the wind at a distance $R_{\rm w, TS}$ and a forward shock propagating in the ISM at $R_{\rm w, FS}$ (Figure \ref{fig:sketch}). These features are located at \citep[e.g.][]{Koo:ApJ:1992},
\begin{align}
   R_{\rm w, TS}&=
0.19{\lw^{3/10}t_3^{2/5}\over \nam^{3/10}\betaw^{1/2}}~{\rm pc};\label{eq:R_TS}\\
  R_{\rm w, FS}&=
  1.8{\lw^{1/5}t_3^{3/5}\over\nam^{1/5}}~{\rm pc}.\label{eq:R_FS}
\end{align}
The typical duration of the super-Eddington phase is $t\sim 10-10^3$~kyr \citep[e.g.][]{Wiktorowicz:ApJ:2017,Shao:ApJ:2020}.
Particles can be accelerated at the wind termination shock, and transported into the nebulae.
The wind nebula expands at a speed of $\dot{R}_{\rm w, FS}\approx 10^3\lw^{1/5}t_3^{-2/5}\nam^{-1/5} ~{\rm km/s}$.
For a longer active time or low density ISM, this can account for observed nebulae with sizes up to hundreds of parsec. 
For an inhomogeneous surrounding medium, the nebula will likely deviate from spherical symmetry.

\bibliography{main}{}

\begin{thebibliography}{}
\expandafter\ifx\csname natexlab\endcsname\relax\def\natexlab#1{#1}\fi
\providecommand{\url}[1]{\href{#1}{#1}}
\providecommand{\dodoi}[1]{doi:~\href{http://doi.org/#1}{\nolinkurl{#1}}}
\providecommand{\doeprint}[1]{\href{http://ascl.net/#1}{\nolinkurl{http://ascl.net/#1}}}
\providecommand{\doarXiv}[1]{\href{https://arxiv.org/abs/#1}{\nolinkurl{https://arxiv.org/abs/#1}}}

\bibitem[{M.~G. {Aartsen} {et~al.}(2019){Aartsen}, {Ackermann}, {Adams},
  {Aguilar}, {Ahlers}, {Ahrens}, {Alispach}, {Andeen}, {Anderson}, {Ansseau},
  {Anton}, {Arg{\"u}elles}, {Auffenberg}, {Axani}, {Backes}, {Bagherpour},
  {Bai}, {Barbano}, {Barwick}, {Baum}, \& et~al.}]{IceTop_CR_proton:PhRvD:2019}
{Aartsen}, M.~G., {Ackermann}, M., {Adams}, J., {et~al.} 2019,
  \bibinfo{title}{{Cosmic ray spectrum and composition from PeV to EeV using 3
  years of data from IceTop and IceCube},} PhRvD, 100, 082002,
  \dodoi{10.1103/PhysRevD.100.082002}

\bibitem[{M.~G. Aartsen {et~al.}(2021)Aartsen, Abbasi, Ackermann, Adams,
  Aguilar, Ahlers, Ahrens, Alispach, Allison, Amin,
  {et~al.}}]{aartsen2021icecube}
Aartsen, M.~G., Abbasi, R., Ackermann, M., {et~al.} 2021,
  \bibinfo{title}{IceCube-Gen2: the window to the extreme Universe,} Journal of
  Physics G: Nuclear and Particle Physics, 48, 060501

\bibitem[{R.~U. {Abbasi} {et~al.}(2021){Abbasi}, {Abe}, {Abu-Zayyad}, {Allen},
  {Arai}, {Barcikowski}, {Belz}, {Bergman}, {Blake}, {Cady}, {Cheon}, {Chiba},
  {Chikawa}, {Fujii}, {Fujisue}, {Fujita}, {Fujiwara}, {Fukushima},
  {Fukushima}, {Furlich}, \& et~al.}]{TALE:ApJ:2021}
{Abbasi}, R.~U., {Abe}, M., {Abu-Zayyad}, T., {et~al.} 2021,
  \bibinfo{title}{{The Cosmic-Ray Composition between 2 PeV and 2 EeV Observed
  with the TALE Detector in Monocular Mode},} ApJ, 909, 178,
  \dodoi{10.3847/1538-4357/abdd30}

\bibitem[{A. {Abdul Halim} {et~al.}(2024){Abdul Halim}, {Abreu}, {Aglietta},
  {Allekotte}, {Almeida Cheminant}, {Almela}, {Aloisio}, {Alvarez-Mu{\~n}iz},
  {Ambrosone}, {Ammerman Yebra}, {Anastasi}, {Anchordoqui}, {Andrada}, {Andrade
  Dourado}, {Andringa}, {Apollonio}, {Aramo}, {Ara{\'u}jo Ferreira}, {Arnone},
  {Arteaga Vel{\'a}zquez}, \& et~al.}]{Auger_anisotropy:2024}
{Abdul Halim}, A., {Abreu}, P., {Aglietta}, M., {et~al.} 2024,
  \bibinfo{title}{{Large-scale Cosmic-ray Anisotropies with 19 yr of Data from
  the Pierre Auger Observatory},} ApJ, 976, 48,
  \dodoi{10.3847/1538-4357/ad843b}

\bibitem[{F. {Aharonian} {et~al.}(2019){Aharonian}, {Yang}, \& {de O{\~n}a
  Wilhelmi}}]{Aharonian:NatAs:2019}
{Aharonian}, F., {Yang}, R., \& {de O{\~n}a Wilhelmi}, E. 2019,
  \bibinfo{title}{{Massive stars as major factories of Galactic cosmic rays},}
  NatAs, 3, 561, \dodoi{10.1038/s41550-019-0724-0}

\bibitem[{F.~A. {Aharonian} {et~al.}(2002){Aharonian}, {Belyanin}, {Derishev},
  {Kocharovsky}, \& {Kocharovsky}}]{Aharonian2002}
{Aharonian}, F.~A., {Belyanin}, A.~A., {Derishev}, E.~V., {Kocharovsky}, V.~V.,
  \& {Kocharovsky}, V.~V. 2002, \bibinfo{title}{{Constraints on the extremely
  high-energy cosmic ray accelerators from classical electrodynamics},} \prd,
  66, 023005, \dodoi{10.1103/PhysRevD.66.023005}

\bibitem[{S. {Aiello} {et~al.}(2019){Aiello}, {Akrame}, {Ameli}, {Anassontzis},
  {Andre}, {Androulakis}, {Anghinolfi}, {Anton}, {Ardid}, {Aublin}, {Avgitas},
  {Bagatelas}, {Barbarino}, {Baret}, {Barrios-Mart{\'\i}}, {Belias}, {Berbee},
  {van den Berg}, {Bertin}, {Biagi}, \& et~al.}]{Aiello:APh:2019}
{Aiello}, S., {Akrame}, S.~E., {Ameli}, F., {et~al.} 2019,
  \bibinfo{title}{{Sensitivity of the KM3NeT/ARCA neutrino telescope to
  point-like neutrino sources},} APh, 111, 100,
  \dodoi{10.1016/j.astropartphys.2019.04.002}

\bibitem[{R. Alfaro {et~al.}(2024)Alfaro, Alvarez, Arteaga-Velázquez,
  Avila~Rojas, Ayala~Solares, Babu, Belmont-Moreno, Caballero-Mora, Capistrán,
  Carramiñana, Casanova, Cotti, Cotzomi, Coutiño~de León, De~la Fuente,
  Depaoli, Di~Lalla, Diaz~Hernandez, Dingus, DuVernois, Durocher, Díaz-Vélez,
  Engel, Espinoza, Fan, Fang, Fraija, Fraija, García-González, Garfias,
  Gonzalez~Muñoz, González, Goodman, Groetsch, Harding, Herzog, Hinton,
  Huang, Hueyotl-Zahuantitla, Hüntemeyer, Iriarte, Joshi, Kaufmann, Kieda,
  de~León, Lee, León~Vargas, Linnemann, Longinotti, Luis-Raya, Malone,
  Martinez, Martínez-Castro, Matthews, Miranda-Romagnoli, Morales-Soto,
  Moreno, Mostafá, Nayerhoda, Nellen, Newbold, Nisa, Noriega-Papaqui,
  Olivera-Nieto, Omodei, Osorio, Pérez~Araujo, Pérez-Pérez, Rho,
  Rosa-González, Ruiz-Velasco, Salazar, Salazar-Gallegos, Sandoval, Schneider,
  Serna-Franco, Smith, Son, Springer, Tibolla, Tollefson, Torres,
  Torres-Escobedo, Turner, Ureña-Mena, Varela, Villaseñor, Wang, Watson,
  Willox, Yun-Cárcamo, \& Zhou}]{HAWC_V4641}
Alfaro, R., Alvarez, C., Arteaga-Velázquez, J.~C., {et~al.} 2024,
  \bibinfo{title}{Ultra-high-energy gamma-ray bubble around microquasar {V4641}
  {Sgr},} Nature, 634, 557, \dodoi{10.1038/s41586-024-07995-9}

\bibitem[{ {Astropy Collaboration} {et~al.}(2022){Astropy Collaboration},
  {Price-Whelan}, {Lim}, {Earl}, {Starkman}, {Bradley}, {Shupe}, {Patil},
  {Corrales}, {Brasseur}, {N{\"o}the}, {Donath}, {Tollerud}, {Morris},
  {Ginsburg}, {Vaher}, {Weaver}, {Tocknell}, {Jamieson}, {van Kerkwijk}, \&
  et~al.}]{AstropyCollaborationApJ2022}
{Astropy Collaboration}, {Price-Whelan}, A.~M., {Lim}, P.~L., {et~al.} 2022,
  \bibinfo{title}{{The Astropy Project: Sustaining and Growing a
  Community-oriented Open-source Project and the Latest Major Release (v5.0) of
  the Core Package},} ApJ, 935, 167, \dodoi{10.3847/1538-4357/ac7c74}

\bibitem[{A.~M. {Atoyan} {et~al.}(1995){Atoyan}, {Aharonian}, \&
  {V{\"o}lk}}]{Atoyan:PhRvD:1995}
{Atoyan}, A.~M., {Aharonian}, F.~A., \& {V{\"o}lk}, H.~J. 1995,
  \bibinfo{title}{{Electrons and positrons in the galactic cosmic rays},}
  PhRvD, 52, 3265, \dodoi{10.1103/PhysRevD.52.3265}

\bibitem[{W.~I. {Axford}(1994){Axford}}]{Axford:ApJS:1994}
{Axford}, W.~I. 1994, \bibinfo{title}{{The Origins of High-Energy Cosmic
  Rays},} ApJS, 90, 937, \dodoi{10.1086/191928}

\bibitem[{R. {Beck}(2015){Beck}}]{Beck:A&ARv:2015}
{Beck}, R. 2015, \bibinfo{title}{{Magnetic fields in spiral galaxies},} A\&ARv,
  24, 4, \dodoi{10.1007/s00159-015-0084-4}

\bibitem[{M.~C. {Begelman} \& D.~F. {Cioffi}(1989){Begelman} \&
  {Cioffi}}]{Begelman:ApJL:1989}
{Begelman}, M.~C., \& {Cioffi}, D.~F. 1989, \bibinfo{title}{{Overpressured
  Cocoons in Extragalactic Radio Sources},} ApJL, 345, L21,
  \dodoi{10.1086/185542}

\bibitem[{M.~C. {Begelman} {et~al.}(2006){Begelman}, {King}, \&
  {Pringle}}]{Begelman:MNRAS:2006}
{Begelman}, M.~C., {King}, A.~R., \& {Pringle}, J.~E. 2006,
  \bibinfo{title}{{The nature of SS433 and the ultraluminous X-ray sources},}
  MNRAS, 370, 399, \dodoi{10.1111/j.1365-2966.2006.10469.x}

\bibitem[{A.~R. {Bell} {et~al.}(2013){Bell}, {Schure}, {Reville}, \&
  {Giacinti}}]{Bell:MNRAS:2013}
{Bell}, A.~R., {Schure}, K.~M., {Reville}, B., \& {Giacinti}, G. 2013,
  \bibinfo{title}{{Cosmic-ray acceleration and escape from supernova
  remnants},} MNRAS, 431, 415, \dodoi{10.1093/mnras/stt179}

\bibitem[{V.~S. {Berezinskii} {et~al.}(1990){Berezinskii}, {Bulanov}, {Dogiel},
  \& {Ptuskin}}]{Berezinskii}
{Berezinskii}, V.~S., {Bulanov}, S.~V., {Dogiel}, V.~A., \& {Ptuskin}, V.~S.
  1990, {Astrophysics of cosmic rays} ({North-Holland})

\bibitem[{W. {Brinkmann} {et~al.}(2005){Brinkmann}, {Kotani}, \&
  {Kawai}}]{SS433_abundance:A&A:2005}
{Brinkmann}, W., {Kotani}, T., \& {Kawai}, N. 2005, \bibinfo{title}{{XMM-Newton
  observations of SS 433 I. EPIC spectral analysis},} A\&A, 431, 575,
  \dodoi{10.1051/0004-6361:20041768}

\bibitem[{Z. {Cao} {et~al.}(2024){Cao}, {Aharonian}, \& {An}}]{LHAASO-catalog2}
{Cao}, Z., {Aharonian}, F., \& {An}, Q., e. 2024, \bibinfo{title}{{The First
  LHAASO Catalog of Gamma-Ray Sources},} \apjs, 271, 25,
  \dodoi{10.3847/1538-4365/acfd29}

\bibitem[{ {Cherenkov Telescope Array Consortium} {et~al.}(2023){Cherenkov
  Telescope Array Consortium}, {Acero}, {Acharyya}, {Adam}, {Aguasca-Cabot}, \&
  et~al.}]{CTA2023APh}
{Cherenkov Telescope Array Consortium}, {Acero}, F., {Acharyya}, A., {et~al.}
  2023, \bibinfo{title}{{Sensitivity of the Cherenkov Telescope Array to
  spectral signatures of hadronic PeVatrons with application to Galactic
  Supernova Remnants},} Astroparticle Physics, 150, 102850,
  \dodoi{10.1016/j.astropartphys.2023.102850}

\bibitem[{E.~M. {Churazov} {et~al.}(2024){Churazov}, {Khabibullin}, \&
  {Bykov}}]{Churazov:A&A:2024}
{Churazov}, E.~M., {Khabibullin}, I.~I., \& {Bykov}, A.~M. 2024,
  \bibinfo{title}{{Minimalist model of the W50/SS433 extended X-ray jet:
  Anisotropic wind with recollimation shocks},} A\&A, 688, A4,
  \dodoi{10.1051/0004-6361/202449343}

\bibitem[{A.~J. {Cooper} {et~al.}(2020){Cooper}, {Gaggero}, {Markoff}, \&
  {Zhang}}]{Cooper:MNRAS:2020}
{Cooper}, A.~J., {Gaggero}, D., {Markoff}, S., \& {Zhang}, S. 2020,
  \bibinfo{title}{{High-energy cosmic ray production in X-ray binary jets},}
  MNRAS, 493, 3212, \dodoi{10.1093/mnras/staa373}

\bibitem[{M. {Coriat} {et~al.}(2019){Coriat}, {Fender}, {Tasse}, {Smirnov},
  {Tzioumis}, \& {Broderick}}]{CirX1jet}
{Coriat}, M., {Fender}, R.~P., {Tasse}, C., {et~al.} 2019, \bibinfo{title}{{The
  twisted jets of Circinus X-1},} \mnras, 484, 1672,
  \dodoi{10.1093/mnras/stz099}

\bibitem[{P. {Cristofari} {et~al.}(2020){Cristofari}, {Blasi}, \&
  {Amato}}]{Cristofari}
{Cristofari}, P., {Blasi}, P., \& {Amato}, E. 2020, \bibinfo{title}{{The low
  rate of Galactic pevatrons},} Astroparticle Physics, 123, 102492,
  \dodoi{10.1016/j.astropartphys.2020.102492}

\bibitem[{S. {Fabrika}(2004){Fabrika}}]{Fabrika:ASPRv:2004}
{Fabrika}, S. 2004, \bibinfo{title}{{The jets and supercritical accretion disk
  in SS433},} ASPRv, 12, 1, \dodoi{10.48550/arXiv.astro-ph/0603390}

\bibitem[{R. {Fender} \& T. {Belloni}(2004){Fender} \&
  {Belloni}}]{Fender:ARA&A:2004}
{Fender}, R., \& {Belloni}, T. 2004, \bibinfo{title}{{GRS 1915+105 and the
  Disc-Jet Coupling in Accreting Black Hole Systems},} ARA\&A, 42, 317,
  \dodoi{10.1146/annurev.astro.42.053102.134031}

\bibitem[{G. {Giacinti} {et~al.}(2015){Giacinti}, {Kachelrie{\ss}}, \&
  {Semikoz}}]{Giacinti2015PhRvD}
{Giacinti}, G., {Kachelrie{\ss}}, M., \& {Semikoz}, D.~V. 2015,
  \bibinfo{title}{{Escape model for Galactic cosmic rays and an early
  extragalactic transition},} \prd, 91, 083009,
  \dodoi{10.1103/PhysRevD.91.083009}

\bibitem[{V.~L. {Ginzburg} \& S.~I. {Syrovatskii}(1964){Ginzburg} \&
  {Syrovatskii}}]{Ginzburg}
{Ginzburg}, V.~L., \& {Syrovatskii}, S.~I. 1964, {The Origin of Cosmic Rays}
  (PERGAMON PRESS)

\bibitem[{S. {Golenetskii} {et~al.}(2003){Golenetskii}, {Aptekar}, {Frederiks},
  {Mazets}, {Palshin}, {Hurley}, {Cline}, \& {Stern}}]{CygX1flare}
{Golenetskii}, S., {Aptekar}, R., {Frederiks}, D., {et~al.} 2003,
  \bibinfo{title}{{Observations of Giant Outbursts from Cygnus X-1},} \apj,
  596, 1113, \dodoi{10.1086/378190}

\bibitem[{R.~J. {Gould} \& G. {Schr{\'e}der}(1966){Gould} \&
  {Schr{\'e}der}}]{Gould66}
{Gould}, R.~J., \& {Schr{\'e}der}, G. 1966, \bibinfo{title}{{Opacity of the
  Universe to High-Energy Photons},} \prl, 16, 252,
  \dodoi{10.1103/PhysRevLett.16.252}

\bibitem[{C.~R. Harris {et~al.}(2020)Harris, Millman, van~der Walt, Gommers,
  Virtanen, Cournapeau, Wieser, Taylor, Berg, Smith, Kern, Picus, Hoyer, van
  Kerkwijk, Brett, Haldane, del R{\'{i}}o, Wiebe, Peterson,
  G{\'{e}}rard-Marchant, Sheppard, Reddy, Weckesser, Abbasi, Gohlke, \&
  Oliphant}]{Numpy}
Harris, C.~R., Millman, K.~J., van~der Walt, S.~J., {et~al.} 2020,
  \bibinfo{title}{Array programming with {NumPy},} Nature, 585, 357,
  \dodoi{10.1038/s41586-020-2649-2}

\bibitem[{H. {He} \&  {LHAASO Collaboration}(2018){He} \& {LHAASO
  Collaboration}}]{he18}
{He}, H., \& {LHAASO Collaboration}. 2018, \bibinfo{title}{{Design of the
  LHAASO detectors},} Radiation Detection Technology and Methods, 2, 7,
  \dodoi{https://doi.org/10.1007/s41605-018-0037-3}

\bibitem[{S. {Heinz} \& R. {Sunyaev}(2002){Heinz} \&
  {Sunyaev}}]{Heinz:A&A:2002}
{Heinz}, S., \& {Sunyaev}, R. 2002, \bibinfo{title}{{Cosmic rays from
  microquasars: A narrow component to the CR spectrum?},} A\&A, 390, 751,
  \dodoi{10.1051/0004-6361:20020615}

\bibitem[{ {H.E.S.S. Collaboration} {et~al.}(2016){H.E.S.S. Collaboration},
  {Abramowski}, {Aharonian}, {Benkhali}, {Akhperjanian}, {Ang{\"u}ner},
  {Backes}, {Balzer}, {Becherini}, {Tjus}, {Berge}, {Bernhard}, {Bernl{\"o}hr},
  {Birsin}, {Blackwell}, {B{\"o}ttcher}, {Boisson}, {Bolmont}, {Bordas},
  {Bregeon}, \& et~al.}]{HESS_GC}
{H.E.S.S. Collaboration}, {Abramowski}, A., {Aharonian}, F., {et~al.} 2016,
  \bibinfo{title}{{Acceleration of petaelectronvolt protons in the Galactic
  Centre},} Natur, 531, 476, \dodoi{10.1038/nature17147}

\bibitem[{ {H.E.S.S. Collaboration} {et~al.}(2022){H.E.S.S. Collaboration},
  {Aharonian}, {Ashkar}, {Backes}, {Barbosa Martins}, {Becherini}, {Berge},
  {Bi}, {B{\"o}ttcher}, {de Bony de Lavergne}, {Bradascio}, {Brose}, {Brun},
  {Bulik}, {Burger-Scheidlin}, {Cangemi}, {Caroff}, {Casanova}, {Cerruti},
  {Chand}, {Chandra}, {Chen}, {Chibueze}, {Cristofari}, {Damascene
  Mbarubucyeye}, {Djannati-Ata{\"\i}}, {Ernenwein}, {Feijen}, {Fichet de
  Clairfontaine}, {Fontaine}, {Funk}, {Gabici}, {Gallant}, {Ghafourizadeh},
  {Giavitto}, {Giunti}, {Glawion}, {Glicenstein}, {Goswami}, {Grondin},
  {H{\"a}rer}, {Haupt}, {Hinton}, {H{\"o}rbe}, {Hofmann}, {Holch}, {Holler},
  {Horns}, {Jamrozy}, {Joshi}, {Jung-Richardt}, {Kasai}, {Katarzy{\'n}ski},
  {Katz}, {Kh{\'e}lifi}, {Klu{\'z}niak}, {Komin}, {Kosack}, {Kostunin}, {Kukec
  Mezek}, {Lang}, {Le Stum}, {Lemi{\`e}re}, {Lemoine-Goumard}, {Lenain},
  {Leuschner}, {Lohse}, {Luashvili}, {Lypova}, {Mackey}, {Majumdar},
  {Malyshev}, {Marandon}, {Marchegiani}, {Marcowith}, {Mart{\'\i}-Devesa},
  {Marx}, {Maurin}, {Meyer}, {Mitchell}, {Moderski}, {Mohrmann}, {Montanari},
  {Moulin}, {Muller}, {Murach}, {Nakashima}, {de Naurois}, {Nayerhoda},
  {Niemiec}, {Ohm}, {Olivera-Nieto}, {de Ona Wilhelmi}, {Ostrowski}, {Panny},
  {Panter}, {Parsons}, {Peron}, {Prokhorov}, {P{\"u}hlhofer}, {Punch},
  {Quirrenbach}, {Rauth}, {Reichherzer}, {Reimer}, {Reimer}, {Renaud},
  {Reville}, {Rieger}, {Rowell}, {Rudak}, {Ruiz-Velasco}, {Sahakian},
  {Salzmann}, {Sanchez}, {Santangelo}, {Sasaki}, {Sch{\"u}ssler}, {Schutte},
  {Schwanke}, {Shapopi}, {Specovius}, {Spencer}, {Stawarz}, {Steenkamp},
  {Steinmassl}, {Steppa}, {Sushch}, {Suzuki}, {Takahashi}, {Tanaka}, {Terrier},
  {Thorpe-Morgan}, {Tsirou}, {Tsuji}, {Tuffs}, {Unbehaun}, {van Eldik}, {van
  Soelen}, {Vecchi}, {Veh}, {Venter}, {Vink}, {Wagner}, {White},
  {Wierzcholska}, {Wong}, {Zacharias}, {Zargaryan}, {Zdziarski}, {Zhu},
  {Zouari}, {{\.Z}ywucka}, {Blackwell}, {Braiding}, {Burton}, {Cubuk},
  {Filipovi{\'c}}, {Tothill}, \& {Wong}}]{HESS_WD1}
{H.E.S.S. Collaboration}, {Aharonian}, F., {Ashkar}, H., {et~al.} 2022,
  \bibinfo{title}{{A deep spectromorphological study of the
  {\ensuremath{\gamma}}-ray emission surrounding the young massive stellar
  cluster Westerlund 1},} \aap, 666, A124, \dodoi{10.1051/0004-6361/202244323}

\bibitem[{ {H.E.S.S. Collaboration} {et~al.}(2024){H.E.S.S. Collaboration},
  {Aharonian}, {Ait Benkhali}, {Aschersleben}, {Ashkar}, {Backes}, {Barbosa
  Martins}, {Batzofin}, {Becherini}, {Berge}, {Bernl{\"o}hr}, {Bi},
  {B{\"o}ttcher}, {Boisson}, {Bolmont}, {de Lavergne}, {Borowska},
  {Bouyahiaoui}, {Breuhaus}, {Brose}, \& et~al.}]{HESS_SS433}
{H.E.S.S. Collaboration}, {Aharonian}, F., {Ait Benkhali}, F., {et~al.} 2024,
  \bibinfo{title}{{Acceleration and transport of relativistic electrons in the
  jets of the microquasar SS 433},} Sci, 383, 402,
  \dodoi{10.1126/science.adi2048}

\bibitem[{A.~M. {Hillas}(1984){Hillas}}]{Hillas:ARA&A:1984}
{Hillas}, A.~M. 1984, \bibinfo{title}{{The Origin of Ultra-High-Energy Cosmic
  Rays},} ARA\&A, 22, 425, \dodoi{10.1146/annurev.aa.22.090184.002233}

\bibitem[{J. {Hinton} \&  {SWGO Collaboration}(2022){Hinton} \& {SWGO
  Collaboration}}]{SWGO2022icrc}
{Hinton}, J., \& {SWGO Collaboration}. 2022, in 37th International Cosmic Ray
  Conference, 23, \dodoi{10.22323/1.395.0023}

\bibitem[{R.~M. {Hjellming} \& M.~P. {Rupen}(1995){Hjellming} \&
  {Rupen}}]{Hjellming:Natur:1995}
{Hjellming}, R.~M., \& {Rupen}, M.~P. 1995, \bibinfo{title}{{Episodic ejection
  of relativistic jets by the X-ray transient GRO J1655 - 40},} Natur, 375,
  464, \dodoi{10.1038/375464a0}

\bibitem[{J.~D. Hunter(2007)Hunter}]{Matplotlib}
Hunter, J.~D. 2007, \bibinfo{title}{Matplotlib: A 2D graphics environment,}
  Computing in Science \& Engineering, 9, 90, \dodoi{10.1109/MCSE.2007.55}

\bibitem[{R. {Iaria} {et~al.}(2001){Iaria}, {Di Salvo}, {Burderi}, \&
  {Robba}}]{CirX1}
{Iaria}, R., {Di Salvo}, T., {Burderi}, L., \& {Robba}, N.~R. 2001,
  \bibinfo{title}{{Spectral Evolution of Circinus X-1 along Its Orbit},} \apj,
  561, 321, \dodoi{10.1086/323226}

\bibitem[{Y.-F. {Jiang} {et~al.}(2014){Jiang}, {Stone}, \&
  {Davis}}]{Jiang:ApJ:2014}
{Jiang}, Y.-F., {Stone}, J.~M., \& {Davis}, S.~W. 2014, \bibinfo{title}{{A
  Global Three-dimensional Radiation Magneto-hydrodynamic Simulation of
  Super-Eddington Accretion Disks},} ApJ, 796, 106,
  \dodoi{10.1088/0004-637X/796/2/106}

\bibitem[{C.~R. {Kaiser} {et~al.}(2004){Kaiser}, {Gunn}, {Brocksopp}, \&
  {Sokoloski}}]{GRS1915speed}
{Kaiser}, C.~R., {Gunn}, K.~F., {Brocksopp}, C., \& {Sokoloski}, J.~L. 2004,
  \bibinfo{title}{{Revision of the Properties of the GRS 1915+105 Jets: Clues
  from the Large-Scale Structure},} \apj, 612, 332, \dodoi{10.1086/422466}

\bibitem[{A. {King} {et~al.}(2023){King}, {Lasota}, \&
  {Middleton}}]{King:NewAR:2023}
{King}, A., {Lasota}, J.-P., \& {Middleton}, M. 2023,
  \bibinfo{title}{{Ultraluminous X-ray sources},} NewAR, 96, 101672,
  \dodoi{10.1016/j.newar.2022.101672}

\bibitem[{S.~S. {Komissarov} \& S.~A.~E.~G. {Falle}(1998){Komissarov} \&
  {Falle}}]{Komissarov:MNRAS:1998}
{Komissarov}, S.~S., \& {Falle}, S.~A.~E.~G. 1998, \bibinfo{title}{{The
  large-scale structure of FR-II radio sources},} MNRAS, 297, 1087,
  \dodoi{10.1046/j.1365-8711.1998.01547.x}

\bibitem[{B.-C. {Koo} \& C.~F. {McKee}(1992){Koo} \& {McKee}}]{Koo:ApJ:1992}
{Koo}, B.-C., \& {McKee}, C.~F. 1992, \bibinfo{title}{{Dynamics of Wind Bubbles
  and Superbubbles. I. Slow Winds and Fast Winds},} ApJ, 388, 93,
  \dodoi{10.1086/171132}

\bibitem[{P.~O. {Lagage} \& C.~J. {Cesarsky}(1983){Lagage} \&
  {Cesarsky}}]{Lagage:A&A:1983b}
{Lagage}, P.~O., \& {Cesarsky}, C.~J. 1983, \bibinfo{title}{{The maximum energy
  of cosmic rays accelerated by supernova shocks.},} A\&A, 125, 249

\bibitem[{ {LHAASO Collaboration}(2024{\natexlab{a}}){LHAASO
  Collaboration}}]{LHAASO_CygBubble}
{LHAASO Collaboration}. 2024{\natexlab{a}}, \bibinfo{title}{{An
  ultrahigh-energy {\ensuremath{\gamma}} -ray bubble powered by a super
  PeVatron},} Science Bulletin, 69, 449, \dodoi{10.1016/j.scib.2023.12.040}

\bibitem[{ {LHAASO Collaboration}(2024{\natexlab{b}}){LHAASO
  Collaboration}}]{LHASSO_microquasar}
{LHAASO Collaboration}. 2024{\natexlab{b}}, \bibinfo{title}{{Ultrahigh-Energy
  Gamma-ray Emission Associated with Black Hole-Jet Systems},} arXiv e-prints,
  arXiv:2410.08988.
\newblock \doarXiv{2410.08988}

\bibitem[{ {LHAASO Collaboration}(2024{\natexlab{c}}){LHAASO
  Collaboration}}]{LHAASOW43}
{LHAASO Collaboration}. 2024{\natexlab{c}}, \bibinfo{title}{{Observation of the
  $\gamma$-ray Emission from W43 with LHAASO},} arXiv e-prints,
  arXiv:2408.09905, \dodoi{10.48550/arXiv.2408.09905}

\bibitem[{ {LHAASO Collaboration} {et~al.}(2021){LHAASO Collaboration}, {Cao},
  {Aharonian}, {An}, {Axikegu}, {Bai}, {Bai}, {Bao}, {Bastieri}, {Bi}, {Bi},
  {Cai}, {Cai}, {Cao}, {Chang}, {Chang}, {Chen}, {Chen}, {Chen}, {Chen}, \&
  et~al.}]{LhaasoCollaboration:Sci:2021}
{LHAASO Collaboration}, {Cao}, Z., {Aharonian}, F., {et~al.} 2021,
  \bibinfo{title}{{Peta-electron volt gamma-ray emission from the Crab
  Nebula},} Sci, 373, 425, \dodoi{10.1126/science.abg5137}

\bibitem[{M.~A. {Malkov} \& F.~A. {Aharonian}(2019){Malkov} \&
  {Aharonian}}]{Malkov_steepspectra}
{Malkov}, M.~A., \& {Aharonian}, F.~A. 2019, \bibinfo{title}{{Cosmic-ray
  Spectrum Steepening in Supernova Remnants. I. Loss-free Self-similar
  Solution},} \apj, 881, 2, \dodoi{10.3847/1538-4357/ab2c01}

\bibitem[{R.~N. {Manchester} {et~al.}(2005){Manchester}, {Hobbs}, {Teoh}, \&
  {Hobbs}}]{ATNFcatalog}
{Manchester}, R.~N., {Hobbs}, G.~B., {Teoh}, A., \& {Hobbs}, M. 2005,
  \bibinfo{title}{{The Australia Telescope National Facility Pulsar
  Catalogue},} \aj, 129, 1993, \dodoi{10.1086/428488}

\bibitem[{J. {Mart{\'\i}} {et~al.}(2001){Mart{\'\i}}, {Paredes}, \&
  {Peracaula}}]{Marti:A&A:2001}
{Mart{\'\i}}, J., {Paredes}, J.~M., \& {Peracaula}, M. 2001,
  \bibinfo{title}{{Development of a two-sided relativistic jet in Cygnus X-3},}
  A\&A, 375, 476, \dodoi{10.1051/0004-6361:20010907}

\bibitem[{J.-M. {Mart{\'\i}}(2019){Mart{\'\i}}}]{Marti:Galax:2019}
{Mart{\'\i}}, J.-M. 2019, \bibinfo{title}{{Numerical Simulations of Jets from
  Active Galactic Nuclei},} Galax, 7, 24, \dodoi{10.3390/galaxies7010024}

\bibitem[{J.~M. {Mart{\'\i}} {et~al.}(1997){Mart{\'\i}}, {M{\"u}ller}, {Font},
  {Ib{\'a}{\~n}ez}, \& {Marquina}}]{Marti:ApJ:1997}
{Mart{\'\i}}, J.~M., {M{\"u}ller}, E., {Font}, J.~A., {Ib{\'a}{\~n}ez},
  J.~M.~Z., \& {Marquina}, A. 1997, \bibinfo{title}{{Morphology and Dynamics of
  Relativistic Jets},} ApJ, 479, 151, \dodoi{10.1086/303842}

\bibitem[{P.~S. {Medvedev} {et~al.}(2018){Medvedev}, {Khabibullin}, {Sazonov},
  {Churazov}, \& {Tsygankov}}]{SS433_abundance:AstL:2018}
{Medvedev}, P.~S., {Khabibullin}, I.~I., {Sazonov}, S.~Y., {Churazov}, E.~M.,
  \& {Tsygankov}, S.~S. 2018, \bibinfo{title}{{An Upper Limit on Nickel
  Overabundance in the Supercritical Accretion Disk Wind of SS 433 from X-ray
  Spectroscopy},} AstL, 44, 390, \dodoi{10.1134/S1063773718060038}

\bibitem[{S. {Migliari} {et~al.}(2002){Migliari}, {Fender}, \&
  {M{\'e}ndez}}]{Migliari:Sci:2002}
{Migliari}, S., {Fender}, R., \& {M{\'e}ndez}, M. 2002, \bibinfo{title}{{Iron
  Emission Lines from Extended X-ray Jets in SS 433: Reheating of Atomic
  Nuclei},} Sci, 297, 1673, \dodoi{10.1126/science.1073660}

\bibitem[{J.~C.~A. {Miller-Jones} {et~al.}(2004){Miller-Jones}, {Blundell},
  {Rupen}, {Mioduszewski}, {Duffy}, \& {Beasley}}]{Miller-Jones:ApJ:2004}
{Miller-Jones}, J. C.~A., {Blundell}, K.~M., {Rupen}, M.~P., {et~al.} 2004,
  \bibinfo{title}{{Time-sequenced Multi-Radio Frequency Observations of Cygnus
  X-3 in Flare},} ApJ, 600, 368, \dodoi{10.1086/379706}

\bibitem[{S. {Mineo} {et~al.}(2012){Mineo}, {Gilfanov}, \&
  {Sunyaev}}]{Mineo:MNRAS:2012}
{Mineo}, S., {Gilfanov}, M., \& {Sunyaev}, R. 2012, \bibinfo{title}{{X-ray
  emission from star-forming galaxies - I. High-mass X-ray binaries},} MNRAS,
  419, 2095, \dodoi{10.1111/j.1365-2966.2011.19862.x}

\bibitem[{G. {Morlino} {et~al.}(2021){Morlino}, {Blasi}, {Peretti}, \&
  {Cristofari}}]{Morlino:MNRAS:2021}
{Morlino}, G., {Blasi}, P., {Peretti}, E., \& {Cristofari}, P. 2021,
  \bibinfo{title}{{Particle acceleration in winds of star clusters},} MNRAS,
  504, 6096, \dodoi{10.1093/mnras/stab690}

\bibitem[{P. {Mukhopadhyay} {et~al.}(2023){Mukhopadhyay}, {Peretti}, {Globus},
  {Simeon}, \& {Blandford}}]{Mukhopadhyay:ApJ:2023}
{Mukhopadhyay}, P., {Peretti}, E., {Globus}, N., {Simeon}, P., \& {Blandford},
  R. 2023, \bibinfo{title}{{Reacceleration of Galactic Cosmic Rays beyond the
  Knee at the Termination Shock of a Cosmic-Ray-driven Galactic Wind},} ApJ,
  953, 49, \dodoi{10.3847/1538-4357/acdc9b}

\bibitem[{J. {Neilsen} {et~al.}(2016){Neilsen}, {Rahoui}, {Homan}, \&
  {Buxton}}]{Neilsen:ApJ:2016}
{Neilsen}, J., {Rahoui}, F., {Homan}, J., \& {Buxton}, M. 2016,
  \bibinfo{title}{{A Super-Eddington, Compton-thick Wind in GRO J1655-40?},}
  ApJ, 822, 20, \dodoi{10.3847/0004-637X/822/1/20}

\bibitem[{Y. {Ohira}(2024){Ohira}}]{Ohira:arXiv:2024}
{Ohira}, Y. 2024, \bibinfo{title}{{Very-high-energy gamma rays from cosmic rays
  escaping from Galactic black hole binaries},} arXiv, arXiv:2410.22976,
  \dodoi{10.48550/arXiv.2410.22976}

\bibitem[{K. {Ohsuga} {et~al.}(2005){Ohsuga}, {Mori}, {Nakamoto}, \&
  {Mineshige}}]{Ohsuga:ApJ:2005}
{Ohsuga}, K., {Mori}, M., {Nakamoto}, T., \& {Mineshige}, S. 2005,
  \bibinfo{title}{{Supercritical Accretion Flows around Black Holes:
  Two-dimensional, Radiation Pressure-dominated Disks with Photon Trapping},}
  ApJ, 628, 368, \dodoi{10.1086/430728}

\bibitem[{J.~A. {Orosz} {et~al.}(2001){Orosz}, {Kuulkers}, {van der Klis},
  {McClintock}, {Garcia}, {Callanan}, {Bailyn}, {Jain}, \&
  {Remillard}}]{V4641_abundance:ApJ:2001}
{Orosz}, J.~A., {Kuulkers}, E., {van der Klis}, M., {et~al.} 2001,
  \bibinfo{title}{{A Black Hole in the Superluminal Source SAX J1819.3-2525
  (V4641 Sgr)},} ApJ, 555, 489, \dodoi{10.1086/321442}

\bibitem[{E. {Peretti} {et~al.}(2025){Peretti}, {Petropoulou}, {Vasilopoulos},
  \& {Gabici}}]{Peretti:A&A:2025}
{Peretti}, E., {Petropoulou}, M., {Vasilopoulos}, G., \& {Gabici}, S. 2025,
  \bibinfo{title}{{Particle acceleration and multi-messenger radiation from
  ultra-luminous X-ray sources: A new class of Galactic PeVatrons},} A\&A, 698,
  A188, \dodoi{10.1051/0004-6361/202452987}

\bibitem[{M. {Perucho} \& J.~M. {Mart{\'\i}}(2007){Perucho} \&
  {Mart{\'\i}}}]{Perucho:MNRAS:2007}
{Perucho}, M., \& {Mart{\'\i}}, J.~M. 2007, \bibinfo{title}{{A numerical
  simulation of the evolution and fate of a Fanaroff-Riley type I jet. The case
  of 3C 31},} MNRAS, 382, 526, \dodoi{10.1111/j.1365-2966.2007.12454.x}

\bibitem[{P. {Predehl} {et~al.}(2020){Predehl}, {Sunyaev}, {Becker}, {Brunner},
  {Burenin}, {Bykov}, {Cherepashchuk}, {Chugai}, {Churazov}, {Doroshenko},
  {Eismont}, {Freyberg}, {Gilfanov}, {Haberl}, {Khabibullin}, {Krivonos},
  {Maitra}, {Medvedev}, {Merloni}, {Nandra}, {Nazarov}, {Pavlinsky}, {Ponti},
  {Sanders}, {Sasaki}, {Sazonov}, {Strong}, \& {Wilms}}]{eROSITA_bubble}
{Predehl}, P., {Sunyaev}, R.~A., {Becker}, W., {et~al.} 2020,
  \bibinfo{title}{{Detection of large-scale X-ray bubbles in the Milky Way
  halo},} Natur, 588, 227, \dodoi{10.1038/s41586-020-2979-0}

\bibitem[{D. {Proga} {et~al.}(2000){Proga}, {Stone}, \&
  {Kallman}}]{Proga:ApJ:2000}
{Proga}, D., {Stone}, J.~M., \& {Kallman}, T.~R. 2000,
  \bibinfo{title}{{Dynamics of Line-driven Disk Winds in Active Galactic
  Nuclei},} ApJ, 543, 686, \dodoi{10.1086/317154}

\bibitem[{M. {Revnivtsev} {et~al.}(2002){Revnivtsev}, {Sunyaev}, {Gilfanov}, \&
  {Churazov}}]{Revnivtsev:A&A:2002}
{Revnivtsev}, M., {Sunyaev}, R., {Gilfanov}, M., \& {Churazov}, E. 2002,
  \bibinfo{title}{{V4641Sgr - A super-Eddington source enshrouded by an
  extended envelope},} A\&A, 385, 904, \dodoi{10.1051/0004-6361:20020189}

\bibitem[{P. {Rossi} {et~al.}(2008){Rossi}, {Mignone}, {Bodo}, {Massaglia}, \&
  {Ferrari}}]{Rossi:A&A:2008}
{Rossi}, P., {Mignone}, A., {Bodo}, G., {Massaglia}, S., \& {Ferrari}, A. 2008,
  \bibinfo{title}{{Formation of dynamical structures in relativistic jets: the
  FRI case},} A\&A, 488, 795, \dodoi{10.1051/0004-6361:200809687}

\bibitem[{L. {Scheck} {et~al.}(2002){Scheck}, {Aloy}, {Mart{\'\i}},
  {G{\'o}mez}, \& {M{\"u}ller}}]{Scheck:MNRAS:2002}
{Scheck}, L., {Aloy}, M.~A., {Mart{\'\i}}, J.~M., {G{\'o}mez}, J.~L., \&
  {M{\"u}ller}, E. 2002, \bibinfo{title}{{Does the plasma composition affect
  the long-term evolution of relativistic jets?},} MNRAS, 331, 615,
  \dodoi{10.1046/j.1365-8711.2002.05210.x}

\bibitem[{Y. {Shao} \& X.-D. {Li}(2020){Shao} \& {Li}}]{Shao:ApJ:2020}
{Shao}, Y., \& {Li}, X.-D. 2020, \bibinfo{title}{{Population Synthesis of Black
  Hole X-Ray Binaries},} ApJ, 898, 143, \dodoi{10.3847/1538-4357/aba118}

\bibitem[{D.~A. {Smith} {et~al.}(2023){Smith}, {Abdollahi}, {Ajello}, {Bailes},
  {Baldini}, {Ballet}, {Baring}, {Bassa}, {Gonzalez}, {Bellazzini}, {Berretta},
  {Bhattacharyya}, {Bissaldi}, {Bonino}, {Bottacini}, {Bregeon}, {Bruel},
  {Burgay}, {Burnett}, {Cameron}, \& et~al.}]{Fermi_pulsar_2023}
{Smith}, D.~A., {Abdollahi}, S., {Ajello}, M., {et~al.} 2023,
  \bibinfo{title}{{The Third Fermi Large Area Telescope Catalog of Gamma-Ray
  Pulsars},} ApJ, 958, 191, \dodoi{10.3847/1538-4357/acee67}

\bibitem[{A.~M. {Stirling} {et~al.}(2001){Stirling}, {Spencer}, {de la Force},
  {Garrett}, {Fender}, \& {Ogley}}]{CygX1flarejet}
{Stirling}, A.~M., {Spencer}, R.~E., {de la Force}, C.~J., {et~al.} 2001,
  \bibinfo{title}{{A relativistic jet from Cygnus X-1 in the low/hard X-ray
  state},} \mnras, 327, 1273, \dodoi{10.1046/j.1365-8711.2001.04821.x}

\bibitem[{A.~W. {Strong} {et~al.}(2007){Strong}, {Moskalenko}, \&
  {Ptuskin}}]{Strong:ARNPS:2007}
{Strong}, A.~W., {Moskalenko}, I.~V., \& {Ptuskin}, V.~S. 2007,
  \bibinfo{title}{{Cosmic-Ray Propagation and Interactions in the Galaxy},}
  ARNPS, 57, 285, \dodoi{10.1146/annurev.nucl.57.090506.123011}

\bibitem[{M. {Su} {et~al.}(2010){Su}, {Slatyer}, \&
  {Finkbeiner}}]{Fermi_bubble}
{Su}, M., {Slatyer}, T.~R., \& {Finkbeiner}, D.~P. 2010, \bibinfo{title}{{Giant
  Gamma-ray Bubbles from Fermi-LAT: Active Galactic Nucleus Activity or Bipolar
  Galactic Wind?},} \apj, 724, 1044, \dodoi{10.1088/0004-637X/724/2/1044}

\bibitem[{A.~J. {Tetarenko} {et~al.}(2017){Tetarenko}, {Sivakoff},
  {Miller-Jones}, {Rosolowsky}, {Petitpas}, {Gurwell}, {Wouterloot}, {Fender},
  {Heinz}, {Maitra}, {Markoff}, {Migliari}, {Rupen}, {Rushton}, {Russell},
  {Russell}, \& {Sarazin}}]{Tetarenko:MNRAS:2017}
{Tetarenko}, A.~J., {Sivakoff}, G.~R., {Miller-Jones}, J.~C.~A., {et~al.} 2017,
  \bibinfo{title}{{Extreme jet ejections from the black hole X-ray binary V404
  Cygni},} MNRAS, 469, 3141, \dodoi{10.1093/mnras/stx1048}

\bibitem[{ {The LHAASO Collaboration} {et~al.}(2025){The LHAASO Collaboration},
  {Cao}, {Aharonian}, {Bai}, {Bao}, {Bastieri}, {Bi}, {Bi}, {Bian}, {Bukevich},
  {Cai}, {Cao}, {Cao}, {Chang}, {Chang}, {Chen}, {Chen}, {Chen}, {Chen},
  {Chen}, \& et~al.}]{TheLHAASOCollaboration:arXiv:2025}
{The LHAASO Collaboration}, {Cao}, Z., {Aharonian}, F., {et~al.} 2025,
  \bibinfo{title}{{First Identification and Precise Spectral Measurement of the
  Proton Component in the Cosmic-Ray `Knee'},} arXiv, arXiv:2505.14447,
  \dodoi{10.48550/arXiv.2505.14447}

\bibitem[{S. {Thoudam} {et~al.}(2016){Thoudam}, {Rachen}, {van Vliet},
  {Achterberg}, {Buitink}, {Falcke}, \& {H{\"o}randel}}]{Thoudam:A&A:2016}
{Thoudam}, S., {Rachen}, J.~P., {van Vliet}, A., {et~al.} 2016,
  \bibinfo{title}{{Cosmic-ray energy spectrum and composition up to the ankle:
  the case for a second Galactic component},} A\&A, 595, A33,
  \dodoi{10.1051/0004-6361/201628894}

\bibitem[{J. {van den Eijnden} {et~al.}(2018){van den Eijnden}, {Degenaar},
  {Russell}, {Wijnands}, {Miller-Jones}, {Sivakoff}, \& {Hern{\'a}ndez
  Santisteban}}]{vandenEijnden:Natur:2018}
{van den Eijnden}, J., {Degenaar}, N., {Russell}, T.~D., {et~al.} 2018,
  \bibinfo{title}{{An evolving jet from a strongly magnetized accreting X-ray
  pulsar},} Natur, 562, 233, \dodoi{10.1038/s41586-018-0524-1}

\bibitem[{J. {van den Eijnden} {et~al.}(2019){van den Eijnden}, {Degenaar},
  {Schulz}, {Nowak}, {Wijnands}, {Russell}, {Hern{\'a}ndez Santisteban},
  {Bahramian}, {Maccarone}, {Kennea}, \& {Heinke}}]{vandenEijnden:MNRAS:2019}
{van den Eijnden}, J., {Degenaar}, N., {Schulz}, N.~S., {et~al.} 2019,
  \bibinfo{title}{{Chandra reveals a possible ultrafast outflow in the
  super-Eddington Be/X-ray binary Swift J0243.6+6124},} MNRAS, 487, 4355,
  \dodoi{10.1093/mnras/stz1548}

\bibitem[{A. {Veledina} {et~al.}(2024){Veledina}, {Muleri}, {Poutanen},
  {Podgorn{\'y}}, {Dov{\v{c}}iak}, {Capitanio}, {Churazov}, {De Rosa}, {Di
  Marco}, {Forsblom}, {Kaaret}, {Krawczynski}, {La Monaca}, {Loktev},
  {Lutovinov}, {Molkov}, {Mushtukov}, {Ratheesh}, {Rodriguez Cavero},
  {Steiner}, \& et~al.}]{Veledina:NatAs:2024}
{Veledina}, A., {Muleri}, F., {Poutanen}, J., {et~al.} 2024,
  \bibinfo{title}{{Cygnus X-3 revealed as a Galactic ultraluminous X-ray source
  by IXPE},} NatAs.tmp, \dodoi{10.1038/s41550-024-02294-9}

\bibitem[{T. {Vieu} \& B. {Reville}(2023){Vieu} \& {Reville}}]{Vieu:MNRAS:2023}
{Vieu}, T., \& {Reville}, B. 2023, \bibinfo{title}{{Massive star cluster origin
  for the galactic cosmic ray population at very-high energies},} MNRAS, 519,
  136, \dodoi{10.1093/mnras/stac3469}

\bibitem[{T. {Vieu} {et~al.}(2022){Vieu}, {Reville}, \&
  {Aharonian}}]{Vieu:MNRAS:2022}
{Vieu}, T., {Reville}, B., \& {Aharonian}, F. 2022, \bibinfo{title}{{Can
  superbubbles accelerate ultrahigh energy protons?},} MNRAS, 515, 2256,
  \dodoi{10.1093/mnras/stac1901}

\bibitem[{H.~J. {V{\"o}lk} \& V.~N. {Zirakashvili}(2004){V{\"o}lk} \&
  {Zirakashvili}}]{Volk:A&A:2004}
{V{\"o}lk}, H.~J., \& {Zirakashvili}, V.~N. 2004, \bibinfo{title}{{Cosmic ray
  acceleration by spiral shocks in the galactic wind},} A\&A, 417, 807,
  \dodoi{10.1051/0004-6361:20040018}

\bibitem[{G. {Wiktorowicz} {et~al.}(2017){Wiktorowicz}, {Sobolewska}, {Lasota},
  \& {Belczynski}}]{Wiktorowicz:ApJ:2017}
{Wiktorowicz}, G., {Sobolewska}, M., {Lasota}, J.-P., \& {Belczynski}, K. 2017,
  \bibinfo{title}{{The Origin of the Ultraluminous X-Ray Sources},} ApJ, 846,
  17, \dodoi{10.3847/1538-4357/aa821d}

\bibitem[{Z. Ye {et~al.}(2023)Ye, Hu, Tian, Chang, Chang, Cheng, Gao, Ge, Gong,
  Guo, {et~al.}}]{ye2023trident}
Ye, Z., Hu, F., Tian, W., {et~al.} 2023, \bibinfo{title}{A
  multi-cubic-kilometre neutrino telescope in the western Pacific Ocean,}
  Nature Astronomy, 7, 1497

\bibitem[{V.~N. {Zirakashvili} \& V.~S. {Ptuskin}(2008){Zirakashvili} \&
  {Ptuskin}}]{Zirakashvili:ApJ:2008}
{Zirakashvili}, V.~N., \& {Ptuskin}, V.~S. 2008, \bibinfo{title}{{Diffusive
  Shock Acceleration with Magnetic Amplification by Nonresonant Streaming
  Instability in Supernova Remnants},} \apj, 678, 939, \dodoi{10.1086/529580}

\bibitem[{P.~T. {{\.Z}ycki} {et~al.}(1999){{\.Z}ycki}, {Done}, \&
  {Smith}}]{Zycki:MNRAS:1999}
{{\.Z}ycki}, P.~T., {Done}, C., \& {Smith}, D.~A. 1999, \bibinfo{title}{{The
  1989 May outburst of the soft X-ray transient GS 2023+338 (V404 Cyg)},}
  MNRAS, 309, 561, \dodoi{10.1046/j.1365-8711.1999.02885.x}

\end{thebibliography}
\bibliographystyle{aasjournal}


\end{CJK*}
\end{document}